\documentclass[iop]{emulateapj}
\usepackage{amsmath, amssymb}
\usepackage{amsmath, bm, mathtools, cancel, empheq, ulem, mathrsfs, natbib}
\usepackage{bm, graphicx}
\usepackage[counterclockwise]{rotating}
\usepackage[colorlinks]{hyperref}
\usepackage{multirow}
\usepackage[dvipsnames]{xcolor}
\hypersetup{
	colorlinks = true,
	linkcolor=blue,
	citecolor=blue
}
\usepackage{etoolbox}
\usepackage[title]{appendix}


\newcommand{\MYhref}[3][blue]{\href{#2}{\color{#1}{#3}}}

\newcommand{\ri}{r_{\rm{i}}}
\newcommand{\ro}{r_{\rm{o}}}
\newcommand{\cp}{c_{\rm{p}}}
\newcommand{\pderiv}[2]{\frac{\partial#1}{\partial#2}}
\newcommand{\sn}[2]{#1\times10^{#2}}

\newcommand{\andd}{\text{and}\ \ \ \ \ }

\newcommand{\rhobar}{\overline{\rho}}
\newcommand{\av}[1]{\langle#1\rangle}

\newcommand{\raf}{{\rm{Ra}}_{\rm{F}}}
\newcommand{\e}{\hat{\bm{e}}}
\newcommand{\curl}{\nabla\times}

\newcommand{\prot}{{\rm{P_{rot}}}}
\newcommand{\prm}{{\rm{Pr_m}}}
\newcommand{\rem}{{\rm{Re_m}}}
\newcommand{\define}{\equiv}

\slugcomment{The Astrophysical Journal, \MakeLowercase{000:00 (19pp), accepted February 20, 2020}}

\shorttitle{Bistability in the Global Solar Dynamo}
\shortauthors{Matilsky \& Toomre}

\begin{document}


\title{Exploring Bistability in the Cycles of the Solar Dynamo through Global Simulations}


\author{Loren I. Matilsky\altaffilmark{1} and Juri Toomre}
\affil{JILA \& Department of Astrophysical and Planetary Sciences, University of Colorado, Boulder, CO 80309-0440, USA}

\altaffiltext{1}{loren.matilsky@colorado.edu}


\begin{abstract}
The calling card of solar magnetism is the sunspot cycle, during which sunspots regularly reverse their polarity sense every 11 years. However, a number of more complicated time-dependent behaviors have also been identified. In particular, there are temporal modulations associated with active longitudes and hemispheric asymmetry, when sunspots appear at certain solar longitudes or else in one hemisphere preferentially. So far, a direct link between between this asymmetric temporal behavior and the underlying solar dynamo has remained elusive. In this work, we present results from global, 3D magnetohydrodynamic (MHD) simulations, which display both behavior reminiscent of the sunspot cycle (regular polarity reversals and equatorward migration of internal magnetic field) and asymmetric, irregular behavior that in the simulations we interpret as active longitudes and hemispheric asymmetry. The simulations are thus bistable, in that the turbulent convection can stably support two distinct flavors of magnetism at different times, in superposition, or with smooth transitions from one state to the other. We discuss this new family of dynamo models in the context of the extensive observations of the Sun's surface magnetic field with the \textit{Solar and Heliospheric Observatory} (\textit{SOHO}) and the \textit{Solar Dynamics Observatory} (\textit{SDO}), as well as earlier observations of sunspot number and synoptic maps. We suggest that the solar dynamo itself may be bistable in nature, exhibiting two types of temporal behavior in the magnetic field.
\end{abstract}




\section{Introduction}
Since the early observations of sunspot number, the Sun's magnetic field has been known to follow the fairly regular 11-year sunspot cycle. Sunspots appear at mid-latitudes at the beginning of the cycle, then at latitudes a bit higher during the peak of solar activity, and finally at sites progressively closer to the equator as the magnetic activity wanes (e.g., \citealt{Hathaway11}). Polarimetric observations from the NSO/Kitt Peak Observatory and magnetograms from the Michelson Doppler Imager (MDI) aboard \textit{SOHO} and the Helioseismic and Magnetic Imager (HMI) aboard \textit{SDO} have enabled detailed and continuous study of how sunspots appear in pairs of opposite polarity sense from one cycle to the next, making a 22-year cycle overall (e.g., \citealt{Hathaway15}). In addition, a number of other cycles (less obvious and regular than the 22-year cycle) have been identified, from a slow $\sim$$100$-year modulation of the peak cycle amplitudes called the Gleissberg Cycle (e.g., \citealt{Gleissberg39}; \citealt{Ogurtsov02}), to a rapid $\sim$2-year periodicity in the global magnetic field (e.g., \citealt{Ulrich13}). On timescales in  between, \textit{active longitudes} (longitudes at which sunspots appear more frequently and with greater strength) persist for several decades (e.g., \citealt{Henney02}). There are also periods (the longest being $\sim$50 yr) associated with \textit{hemispheric asymmetry}, or times at which more sunspots conglomerate in either the northern or southern hemispheres (e.g., \citealt{Ballester05}). Finally, there are several recorded \textit{grand minima}, or times throughout history when solar activity is substantially diminished over protracted intervals of several decades (e.g., \citealt{Hoyt98}).

All of these complicated features of the solar cycle must have their roots in the interior solar dynamo---the process by which the Sun's interior magnetic field regenerates through dynamical interactions between rotation and convection. Sunspots are believed to be the result of rising toroidal loops of magnetic field. That their sites of emergence migrate equatorward with the sunspot cycle and that their polarity senses flip every 11 years suggests an interior toroidal reservoir of magnetism which also migrates equatorward and flips polarity every 11 years. The additional cycles and behavior mentioned above could also be indicative of corresponding cycles in the interior field.

In the past decade, global 3D magnetohydrodynamic (MHD) simulations of the solar convection zone have made significant headway in reproducing aspects of the solar cycle. Using the anelastic spherical harmonic (ASH) code, \citet{Brown10, Brown11} have shown it is possible to build strong interior ``wreaths" of magnetism amidst chaotic, turbulent flow. These wreaths can be space-filling, with nearly 360$^\circ$-connectivity throughout the spherical shell, and can exhibit regular polarity reversals, during which the longitudinal direction of magnetic field in the wreath flips every few years of simulation time. Also using ASH, \citet{Augustson15} have achieved wreath-building dynamos which exhibit both equatorward propagation of wreaths at low latitudes and poleward propagation of wreaths at high latitudes---a signature feature of the \textit{SOHO}/\textit{SDO} observations (e.g., \citealt{Hathaway15}). 

Perhaps one of the most significant issues facing global MHD simulations of solar convection is that the wide range of spatial and temporal scales relevant for the fluid makes a direct numerical simulation impossible. Researchers have attempted to address this problem by using various prescriptions for sub-grid-scale (SGS) turbulent effects or by restricting the global domain. \citet{Warnecke14} used the \textsc{Pencil Code} to solve the MHD equations in a spherical-wedge geometry, showing that global convection could yield $\alpha\Omega$-type dynamos with the direction of propagation of the interior magnetic field being set by the Parker-Yoshimura rule. \citet{Passos14} used the EULAG-MHD code (which incorporates implicit dissipation on the smallest spatial scales to maintain numerical stability) to achieve regular, cyclic polarity reversals in their global \textit{millenium simulation}. The cycles persisted over very long time scales, with statistical features showing long-term trends reminiscent of the observed solar Gleissberg modulation. \citet{Hotta16} have explored a large dynamical range for solar convection using their \textit{reduced speed of sound technique}. They explored the coupling of a near-surface layer of small-scale ($<10$ Mm) convection and deep, large-scale flows, finding that coherent magnetic structures could persist even in the presence of very small diffusivities.
	
Another important outstanding issue for solar MHD simulations is that most lack the eruption of interior magnetic flux and subsequent decay of active regions believed to play an essential role in the global dynamo. In the work of \citet{Nelson11, Nelson14}, an extension of ASH incorporating a Dynamic Smagorinsky treatment of the sub-grid-scale (SGS) fluid motions achieved a dynamo in which small loops of magnetism detached from the interior wreaths and rose to the outer surface via magnetic buoyancy instabilities. The polarity, twist, and tilt of the loops displayed statistical properties reminiscent of Joy's law for sunspot emergence. Using a finite-difference spherical anelastic MHD code, \citet{Fan14} found that the convection gave rise to super-equipartition magnetic flux bundles that had similar characteristics to emerging active regions on the Sun. More recently, dynamos with wreaths that give rise to buoyant loops have been achieved with Rayleigh (the code used in this work) in modeling the intense magnetism  exhibited by M-dwarf stars \citep{Bice20}. These simulations have thus shown it is possible to connect MHD simulations of the interior solar dynamo to the emergence of magnetic flux that is actually observed at the photosphere.

It has generally been difficult for simulations to reproduce equatorward propagation of magnetic field as seen in the solar butterfly diagram (although see \citealt{Kapyla12} and \citealt{Augustson15} for notable exceptions). Furthermore, two other prominent features of solar magnetic activity---active longitudes and hemispheric asymmetry---have not been systematically explored in simulations. In this work, we report on a new class of dynamo simulations whose magnetism exhibits polarity-reversing cycles with equatorward-propagation, as well as a quasi-regular, hemispherically asymmetric cycle, the features of which are suggestive of active longitudes and hemispheric asymmetry. Both cycles are present simultaneously, although usually one cycle is more dominant than the other. 

In Section \ref{sec:num} we discuss our numerical methods and the parameter space covered by our simulations, as well as the hydrodynamic progenitor common to each magnetic simulation. In Section \ref{sec:cycles} we discuss the bistable nature of the magnetism achieved in our dynamo cases. We discuss striking flux-transport-like behavior in Section \ref{sec:polarcaps}. We examine each cycling mode individually in Sections \ref{sec:fourfold} and \ref{sec:partial}. In Sections \ref{sec:cycles}--\ref{sec:partial} we refer only to our primary lowest-magnetic-Prandtl-number case, and we return to the higher-magnetic-Prandtl-number cases in Section \ref{sec:higher_pm}. In Section \ref{sec:dyn} we explore the dynamical origins of the two cycles in terms of the production of magnetic field. In Section \ref{sec:obs} we discuss our bistable dynamo simulations in the context of solar observations and present our conclusions in Section \ref{sec:concl}. 

\section{Numerical Experiment}\label{sec:num}
We consider time-dependent, 3D simulations of a rotating, convecting shell of fluid modeled after the solar convection zone (CZ). We drive the system via a fixed internal heating profile that represents radiative diffusion and injects a solar luminosity (distributed in space) into the layer. The background stratification is marginally stable to convection (i.e., adiabatic), although the heating quickly drives the system to state of instability, wherein convection carries most of the solar luminosity throughout the bulk of the CZ. The energy injected by the heating is ultimately carried out the top boundary via thermal conduction. 

We use the open-source, pseudo-spectral MHD code Rayleigh 0.9.1, which scales efficiently on parallel architectures \citep{Featherstone16a, Matsui16, Featherstone18}. Our modeling is computationally quite challenging and we often used of order 10,000 processors on the Pleiades supercomputer. Rayleigh solves the equations of MHD in rotating, convecting spherical shells, expanding the fluid and magnetic variables with spherical harmonics in the horizontal directions and with Chebyshev polynomials in the radial direction. The spherical shell has inner radius $\ri$ and outer radius $\ro$. In discussing the mathematics, we use the standard spherical coordinates $r$, $\theta$, and $\phi$---the radial distance, co-latitude, and azimuthal angle, respectively---and corresponding unit vectors $\hat{\bm{e}}_r$, $\hat{\bm{e}}_\theta$, and $\hat{\bm{e}}_\phi$. 

Rayleigh implements an anelastic approximation (e.g., \citealt{Gough68}; \citealt{Gilman81}; \citealt{Jones11}), which effectively filters out sound waves, making the maximum allowable timestep limited by the flow velocity and not the sound speed. The thermodynamic variables are linearized about a temporally steady and spherically symmetric reference state with adiabatic stratification, similar to the stratification of the solar CZ revealed by helioseismology. The pressure, density, temperature, and entropy are denoted by $P$, $\rho$, $T$, and $S$, respectively. Overbars indicate the fixed reference state and the absence of overbars indicate deviations from the reference state. Although the reference state is fixed in time, the simulations naturally develop spherically symmetric contributions to the deviations $P$, $\rho$, $T$, and $S$, which effectively allow the system to achieve a final equilibrium state that is slightly different from the reference state. 

\subsection{Fluid equations and boundary conditions}
Under the assumptions of the anelastic approximation, the nonlinear fluid MHD equations are given by (e.g., \citealt{Brown10})
\begin{align}
\nabla\cdot(\overline{\rho}\bm{v}) &=  0,
\label{eq:cont}
\end{align}
\begin{align}
\nabla\cdot\bm{B} &=  0,
\label{eq:divb0}
\end{align}
\begin{align}
\overline{\rho}\Bigg{[}\frac{\partial\bm{v}}{\partial t} + (\bm{v}\cdot\nabla)\bm{v}\Bigg{]} = &-\overline{\rho}\nabla \Bigg{(}\frac{P}{\overline{\rho}}\Bigg{)} -\frac{\overline{\rho} S}{\cp}\bm{g} + \nabla\cdot \bm{D}\nonumber\\
&-2\overline{\rho}\bm{\Omega}_0\times\bm{v} + \frac{1}{4\pi}(\curl\bm{B})\times\bm{B},
\label{eq:mom}
\end{align}
\begin{align}
\overline{\rho}\overline{T}\Bigg{[}\frac{\partial S}{\partial t} + \bm{v}\cdot\nabla S\Bigg{]} =\ &\nabla\cdot\big{[}\kappa\overline{\rho}\overline{T}\nabla S\big{]} + Q + \frac{\eta}{4\pi}|\curl\bm{B}|^2 \nonumber \\
&+ 2\overline{\rho}\nu\Big{[}e_{ij}e_{ij} - \frac{1}{3}(\nabla\cdot\bm{v})^2\Big{]},
\label{eq:en}
\end{align}
and
\begin{align}
\pderiv{\bm{B}}{t}  =\ &\curl [\bm{v}\times\bm{B} - \eta\curl\bm{B}]. 
\label{eq:ind}
\end{align}
Here $\bm{v} = (v_r, v_\theta, v_\phi)$ is the fluid velocity in the rotating frame, $\bm{B} = (B_r, B_\theta, B_\phi)$ is the magnetic field (also in the rotating frame), $\cp$ is the specific heat at constant pressure, and $\bm{g}(r)$ is the gravitational acceleration due to a solar mass $M_\odot$ at the origin. The momentum equation \eqref{eq:mom} employs the Lantz-Braginsky-Roberts approximation \citep{Lantz92, Braginsky95}, which is exact for an adiabatic reference state. We choose the diffusivities $\nu(r)$, $\kappa(r)$, and $\eta(r)$ (kinematic viscosity, thermometric conductivity, and electrical resistivity, respectively) to be temporally steady and spherically symmetric, varying with radius like $\rhobar(r)^{-1/2}$. Because it is not computationally possible to resolve the full range of motion down to the Kolmogorov scale, these diffusivities must be regarded as turbulent ``eddy" diffusivities. For simplicity, we choose the turbulent $\nu(r)$, $\kappa(r)$, and $\eta(r)$ to appear in the equations of MHD with the same form as the molecular diffusivities, but with greatly enhanced values. The tensors $\bm{D}$ and $e_{ij}$ refer to the Newtonian viscous stress and rate-of-strain, respectively. We use the internal heat source $Q(r)$ to represent heating due to radiation, which is chosen to have the fixed radial profile $Q(r)=\alpha[\overline{P}(r) - \overline{P}(\ro)]$. The choice of normalization constant $\alpha$ ensures that the solar luminosity $L_\odot$ is forced through the domain (see \citealt{Featherstone16a}).

The equation of state considers small quasistatic thermodynamic perturbations about a perfect gas:
 \begin{align}
\frac{\rho}{\overline{\rho}} = \frac{P}{\overline{P}} - \frac{T}{\overline{T}} = \frac{P}{\gamma\overline{P}}- \frac{S}{\cp},\label{eq:eos}
\end{align}
$\gamma = 5/3$ being the ratio of specific heats for a fully ionized gas (i.e., a gas with three translational degrees of freedom). 

We adopt stress-free and impenetrable boundary conditions on the velocity:
\begin{align}
v_r = \pderiv{}{r}\bigg{(}\frac{v_\theta}{r}\bigg{)}  = \pderiv{}{r}\bigg{(}\frac{v_\phi}{r}\bigg{)} = 0\ \text{at}\ r=\ri\ \text{and}\ \ro\label{eq:vbc}.
\end{align}
For the magnetic field, we match to a potential field at both boundaries: 
\begin{align}
\bm{B} = \nabla\Phi\ \text{at}\ r=\ri\ \text{and}\ \ro,\ \text{where}\ \nabla^2\Phi = 0. \label{eq:bbc}
\end{align}
Other magnetic lower boundary conditions, such as a perfect conductor or a purely radial field, could be considered, but there is no obvious choice from a physical point of view.

Finally, we force the entropy perturbation $S$ to satisfy a fixed-flux condition at both boundaries:
\begin{align}
 \pderiv{S}{r} &= 0\ \text{at}\ r=\ri \label{eq:sbc_inner}\\
 \andd\pderiv{S}{r} &= -\frac{L_\odot}{4\pi r^2\kappa\rhobar\overline{T}}\ \text{at}\ r=\ro.
 \label{eq:sbc_outer}
 \end{align}
 
  \begin{table}
 	\caption{Input parameter values common to all four cases (H3, D3-1, D3-2, and D3-4).}\label{tab:hydro}
 	\centering
 	\begin{tabular}{r  l}
 		\hline\hline
 		$\ri$       & $\sn{5.00}{10}$ cm\\
 		$\ro$  & $\sn{6.59}{10}$ cm\\
 		$\cp$ & $\sn{3.50}{8}$ erg K$^{-1}$ g$^{-1}$\\
 		$\gamma$  & 5/3 \\
 		$\overline{\rho}_{\rm{i}}$ & 0.181 g cm$^{-3}$ \\
 		$L_\odot$ & $\sn{3.85}{33}$ erg s$^{-1}$\\
 		$M_\odot$ & $\sn{1.99}{33}$ g\\
 		$\Omega_0$ & $\sn{8.61}{-6}$ rad $\rm{s}^{-1}$\\
 		$\prot = 2\pi/\Omega_0$ & 8.45 days\\
 		$\nu(\ro)$ & $\sn{3.00}{12}$ cm$^2$ s$^{-1}$ \\
 		$\rm{Ra}_{\rm{F}}$ & $\sn{2.56}{6}$\\
 		Pr $=  \nu/\kappa$ & 1 \\
 		\hline
 	\end{tabular}
 	\tablecomments{See Appendix \ref{ap:nond} for the definition of the flux Rayleigh number $\raf$.}
 \end{table}

The outer thermal boundary condition \eqref{eq:sbc_outer} causes a sharp radial gradient in the entropy---i.e., a thermal boundary layer---to develop near the outer surface, such that the solar luminosity injected by the internal heating is ultimately carried out via thermal conduction. We note that the microphysics of this thermal boundary layer stand in contrast to the small-scale radiative cooling at the solar photosphere. However, including 3D, non-local radiative transfer in global simulations is currently very difficult computationally and thus beyond the scope of this work. 

\subsection{Parameter Space of the Simulation Suite}
We discuss three dynamo simulations in the present study, which all start from the same hydrodynamic progenitor. We define $\Omega_\odot \define 2.87\times10^{-6}\ \rm{rad}\ \rm{s}^{-1}$ as the sidereal Carrington rotation rate, and all our models rotate at three times this rate to ensure a low Rossby number and thus a solar-like differential rotation, in which the equator rotates faster than the poles (e.g., \citealt{Brown10}). The transition from solar to anti-solar differential rotation as simulations grow increasingly more turbulent has come to be called the ``convective conundrum" \citep{OMara16}. The anti-solar regime can be avoided by either rotating faster or lowering the luminosity, and we opt for the former. We refer to the hydrodynamic progenitor as case H3 (``H" for hydrodynamic, ``3" for $\Omega_0 = 3\Omega_\odot$). Some input parameters common to case H3 and the three dynamo cases are shown in Table \ref{tab:hydro}. 

The three dynamo simulations were initialized from the well-equilibrated hydrodynamic case H3 after about 6400 $\prot$, using the full MHD equations. These dynamo cases differ in the values of the magnetic Prandtl number $\prm = \nu/\eta$, which takes on the values 1, 2, and 4. For all cases, the kinematic viscosity profile $\nu(r)$ is fixed, effectively keeping the Rayleigh number constant. We set the value of $\prm$ by varying the value of the magnetic diffusivity at the top of the domain. We label the corresponding dynamo simulations as cases D3-1, D3-2, and D3-4, where the last number corresponds to the value of $\prm$ being 1, 2, and 4, in turn. To resolve both the flow field and magnetic structures, we use a substantial resolution of at least $(N_r, N_\theta, N_\phi) = (96, 384, 768)$ for the number of radial, latitudinal, and longitudinal grid points, respectively (see Table \ref{tab:nond}). 

Our dynamo cases expand the parameter space explored by \citet{Brown10}---who examined the case $\prm=0.5$ while varying the rotation rate---to higher magnetic Prandtl numbers. In all cases, the magnetic field was initialized from a random seed field of amplitude $\sim$$1$ G (``random" in terms of the amplitude of each individual spectral component).

\begin{figure*}
	\includegraphics{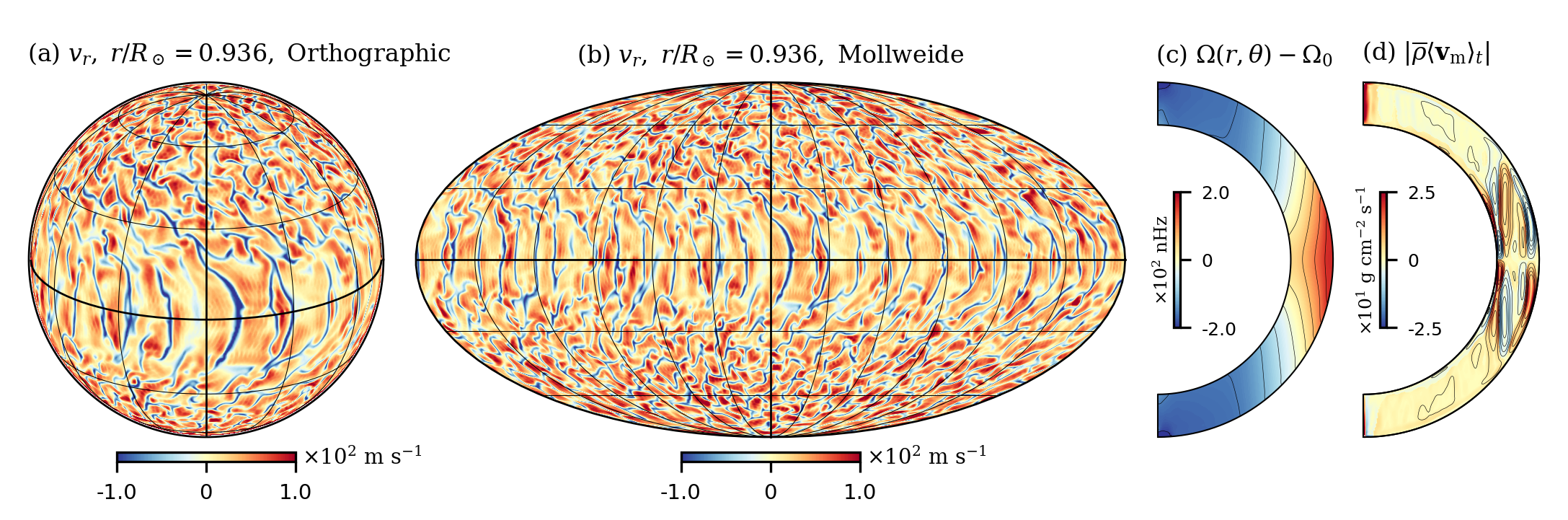}
	\caption{Flow structures in the hydrodynamic case H3. (\textit{a}) Snapshot of the convection (as traced by $v_r$) near the outer surface, shown in orthographic projection. (\textit{b}) The same snapshot of $v_r$, shown in Mollweide projection. (\textit{c}) Averaged fluid rotation rate ($\Omega(r,\theta) - \Omega_0 \equiv \av{v_\phi}_t/r\sin\theta$) in the meridional plane. (\textit{d}) Averaged meridional circulation, with the mass flux magnitude in color overlaid on circulation streamlines. In (\textit{d}), red and blue tones correspond to clockwise and counterclockwise circulation, respectively. For any vector field $\bm{A}$, we define the meridional part as $\bm{A}_{\rm{m}} \equiv A_r\e_r + A_\theta \e_\theta$. The angular brackets with the ``t" subscript denote a combined temporal and longitudinal average.}
	\label{fig:H3}
\end{figure*}

Solutions to the nonlinear MHD equations in rotating, convecting spherical shells involve highly time-dependent and intricate flow structures. Figure \ref{fig:H3} shows a sample of the global flow structure achieved in the hydrodynamic progenitor H3. As seen in the orthographic and Mollweide projections of the radial velocity $v_r$, the flows at any given time consist of Taylor columns between about $\pm20^\circ$-latitude (also called ``Busse columns" and ``banana cells" in the literature) and a broken honeycomb network of upflows and downflows at higher latitudes. The Taylor columns transport angular momentum outward \citep{Brun02, Busse02, Matilsky19} and drive a strong differential rotation with an associated meridional circulation, as seen in Figures \ref{fig:H3}(\textit{c}, \textit{d}). 



 \section{Bistable Magnetic Cycles in the Dynamo Cases}\label{sec:cycles}
The magnetism in our dynamo cases exhibits two striking (and distinct) cycling modes. Usually one is dominant, but they can also coexist simultaneously. The phenomenon of \textit{bistability} in dynamical systems generally refers to the coexistence of two stable equilibrium states, and we use the term somewhat loosely to refer to the two different types of dynamo cycle. In this and the following sections, we discuss case D3-1, which exhibits bistable cycling in the cleanest manner. We return to the higher-$\prm$ cases in Section \ref{sec:higher_pm}. 
 
\begin{figure}
 	\includegraphics{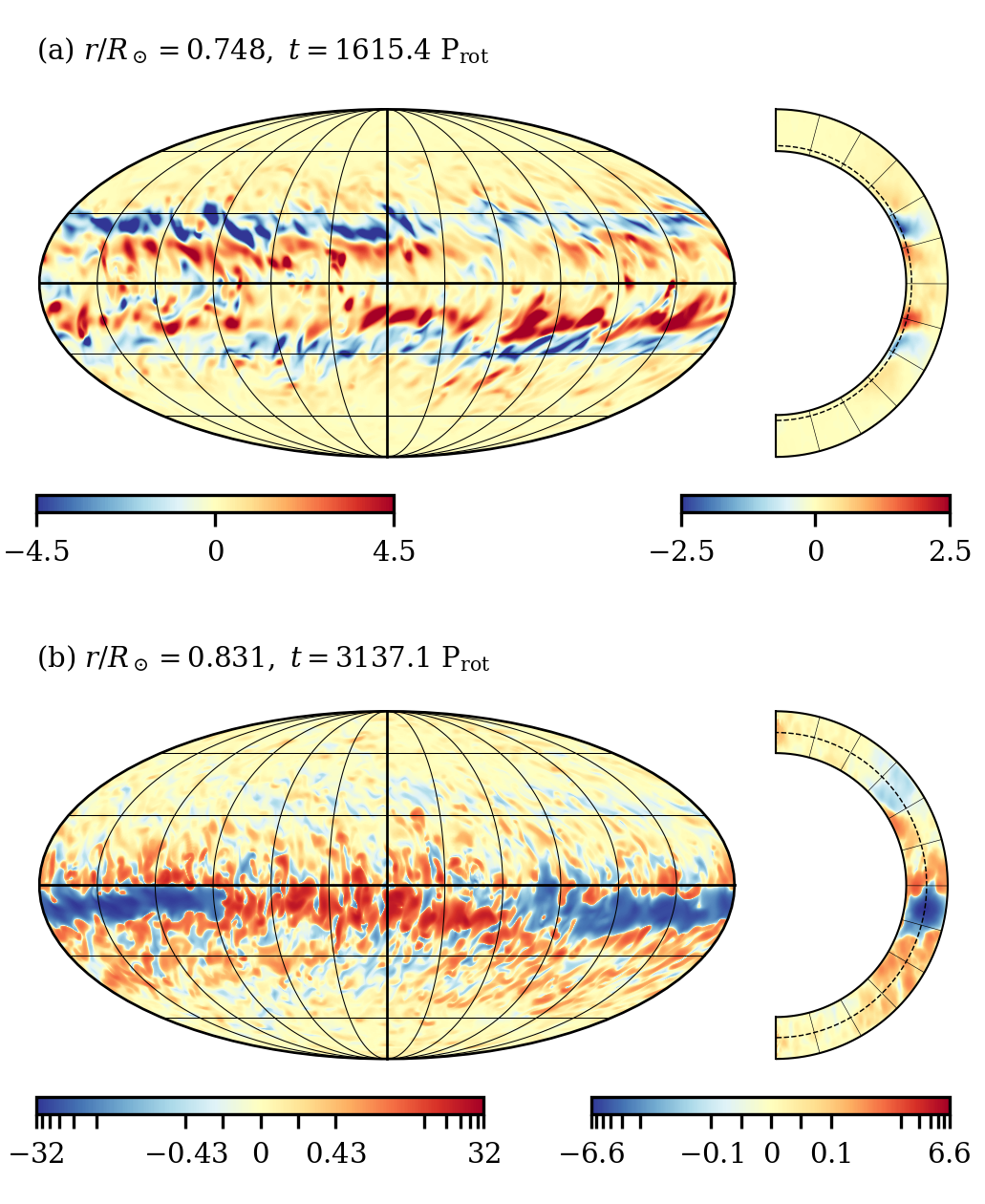}
 	\caption{Snapshots of the azimuthal magnetic field $B_\phi$ (in kG) for case D3-1 at two different depth/time pairs, shown in global Mollweide view on the left and accompanied by the longitudinal average in the meridional plane on the right. (\textit{a}) The field deep in the shell at an early time in the simulation. (\textit{b}) The field in the middle of the shell at a later time. In panel (\textit{b}), the color maps are symmetric-logarithmic to capture both weak-amplitude and strong-amplitude magnetic structures. The colors in the interval around zero map linearly to field strength and the two intervals on either side (one positive and one negative) map logarithmically.  Animated versions of panels (\textit{a}) and (\textit{b}) are available in the online journal.}
 	\label{fig:bphi_twodepths}
 \end{figure}

 Each cycle is associated with a unique morphology of the magnetic field, as seen in Figure \ref{fig:bphi_twodepths}. In roughly the lower half of the shell at an early time as in Figure \ref{fig:bphi_twodepths}(\textit{a}),  there is one pair of opposite-polarity wreaths in each hemisphere (a \textit{fourfold-wreath} structure) confined to within $\sim$$40^\circ$-latitude of the equator. At this instant, the polarity sense of the wreath structure is largely symmetric across the equator. Each wreath mostly has 360$^\circ$-connectivity, linking the magnetic field in a large torus. The wreaths individually have rms field strengths of $\sim$5 kG and longitude-averaged rms field strengths $\sim$2.5 kG. 
 
 Later in the simulation, near the middle of the CZ as in Figure \ref{fig:bphi_twodepths}(\textit{b}), there is a strong, negative-$B_\phi$ \textit{partial wreath} (wrapping $\sim$$180^\circ$ around the sphere), as well as a slightly less prominent positive-$B_\phi$ partial wreath on the opposite side of the sphere. The negative-$B_\phi$ partial wreath is dominant at this time, with a peak strength of $\sim$$40$ kG, compared to the positive-$B_\phi$ peak field strength of $\sim$20 kG. In the longitudinal average, there is a residual $\av{B_\phi}$ that is negative and has a peak magnitude of $\sim$10 kG. The partial-wreath structure occupies most of the domain in radius, but is confined between the equator and $20^\circ$-south in latitude.  
 
\begin{figure*}
	\includegraphics{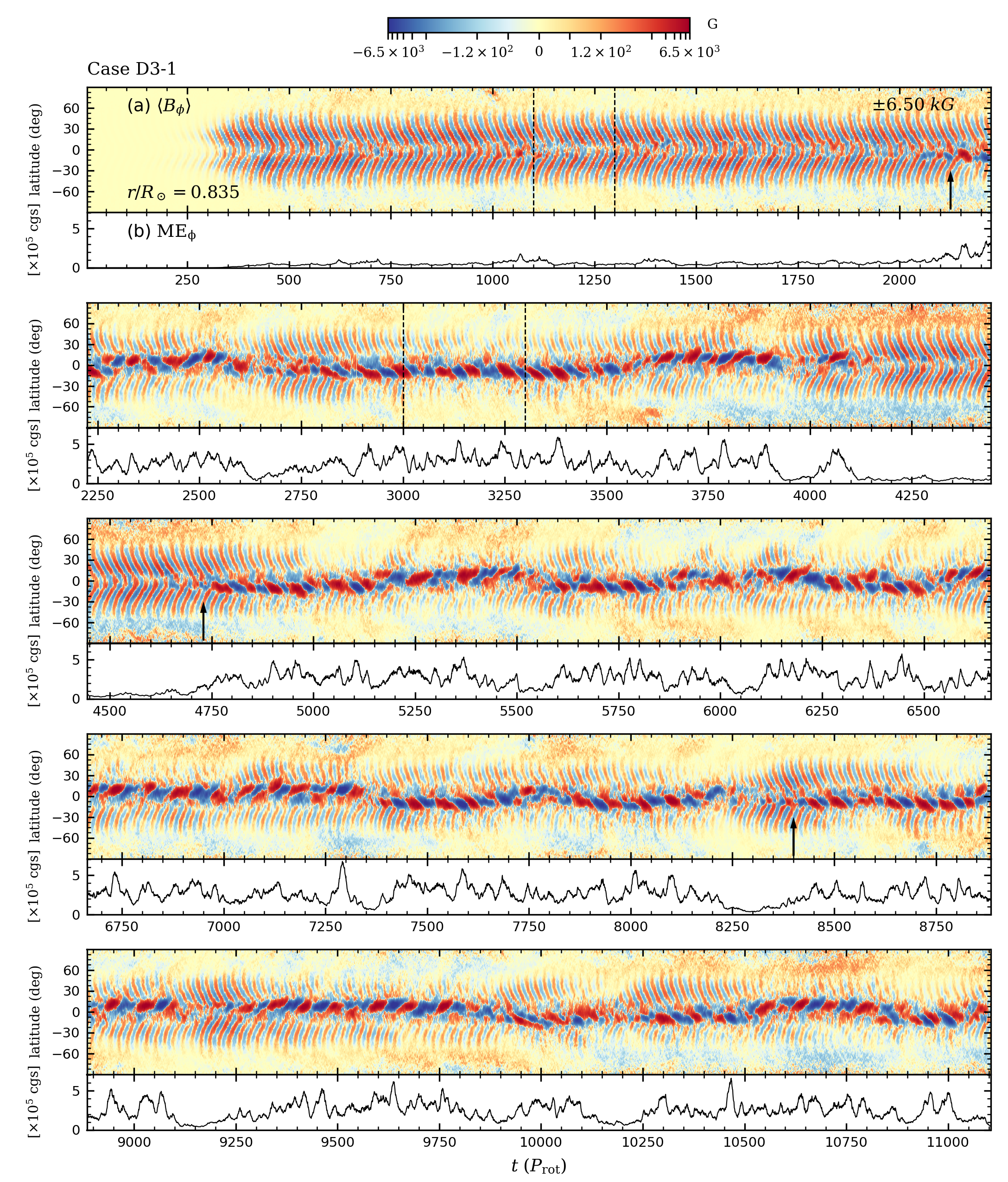}
	\caption{Extended time-latitude diagram of the azimuthal magnetic field $\av{B_\phi}$ for case D3-1 at $r/R_\odot = 0.835$ (mid-CZ). Panels ($a$) (split across multiple rows) show the longitude-averaged field $\av{B_\phi}$ as a function of time and latitude, with the intensity of color corresponding to the magnitude of the field. Time is measured from the instant when the seed magnetic field was introduced to the hydrodynamic progenitor. The saturation values of the color map are indicated in the top row, and the map is symmetric-logarithmic, as in Figure \ref{fig:bphi_twodepths}(\textit{b}). Sharing the same time axis, panels ($b$) show the energy density in the azimuthal magnetic field, ${\rm{ME_\phi}}=B_\phi^2/8\pi$, averaged about $r/R_\odot = 0.835$ over 10\% of the shell by radius. In panels ($b$), the y-axis is measured in the cgs units of energy density, $\rm{erg\ cm^{-3}}$. The arrows mark wreaths in the fourfold-wreath cycle originating at mid-latitudes that appear to seed partial wreaths in the south. The dashed lines mark the intervals shown in closeup view in Figures \ref{fig:zoom_tl_bphi_bot} and \ref{fig:zoom_tl_bphi_mid}.}
	\label{fig:tl_bphi_mid}
\end{figure*}

The temporal behavior of the magnetic field as a whole roughly consists of transitions between the two magnetic field structures presented in Figure \ref{fig:bphi_twodepths}. Figure \ref{fig:tl_bphi_mid}(\textit{a}) shows the evolution in time-latitude space near mid-CZ for $\av{B_\phi}$ (throughout the text, the angular brackets with no subscript denote a longitudinal average at a particular time). After the dynamo has established strong fields from the initial seed field (around $t=300\ \prot$), the \textit{fourfold-wreath cycle} dominates. Each wreath emerges at $\sim$$40^\circ$--$ 50^\circ$-latitude and then migrates steadily equatorward. The newly formed wreaths alternate between positive-$\av{B_\phi}$ and negative-$\av{B_\phi}$ being dominant, with the time between the appearance of two wreaths of the same polarity---the fourfold-wreath cycle period---equal to $\sim$$25$ $\prot$, or about six months. This is the same time it takes an individual wreath to migrate from its mid-latitude starting point to the equator. Effectively each hemisphere operates on its own (i.e., not necessarily in phase), producing wreaths of a given polarity at mid-latitudes that re-appear once they reach the equator. 

At around $t = 2100\ \prot$, the fourfold-wreath cycles are disturbed by the partial-wreath state, which starts in the south (first arrow in Figure \ref{fig:tl_bphi_mid}(\textit{a})). The dominant polarity of the partial-wreath structure reverses quasi-periodically. We can estimate the period (time between two successive states of one dominant polarity) from a visual inspection of Figure \ref{fig:tl_bphi_mid}(\textit{a}): between $t=2950$ $\prot$ and $t=3350$ $\prot$ (an interval slightly larger than the one spanned by the second set of dashed lines), there are roughly five cycles, yielding a period of $\sim$$80$ $\prot$. We note that visual inspections of other intervals would yield slightly different cycle periods, making the period of $\sim$$80$ $\prot$ only approximate. Nevertheless, we thus identify a \textit{partial-wreath cycle} with a period about three times as long as that of the fourfold-wreath cycle. 

The partial-wreath pair wanders into the north around $t = 2250\ \prot$ and flips from north to south once more before the fourfold-wreath cycle becomes dominant for the second time around $t = 4150\ \prot$. The rest of the simulation displays a complex seesaw behavior between the two cycling states. The fourfold-wreath cycle never really disappears, but does significantly decrease in amplitude when the partial-wreath cycle is dominant. Examining Figure \ref{fig:tl_bphi_mid}(\textit{a}) in detail, it seems that whenever the partial-wreath cycle starts, its field comes originally from the fourfold-wreath cycle. That is, a wreath of one polarity will migrate toward the equator and then significantly grow in amplitude, seeding the following partial-wreath cycle with that same polarity dominant. Three instances of this phenomenon are marked by the arrows in Figure \ref{fig:tl_bphi_mid}(\textit{a}) .

Along with the time-latitude panels in Figure \ref{fig:tl_bphi_mid}(\textit{a}), we show the temporal behavior of the energy in the azimuthal magnetic field in Figure \ref{fig:tl_bphi_mid}(\textit{b}), averaged over 10\% of the shell (by radius) at mid-depth. The fourfold-wreath cycle is seen to correspond to a low-energy state of the magnetic field, with sporadic peaks in energy but no discernible energy cycle associated with the polarity reversals.  When the partial-wreath cycle first begins (around $t = 2100\ \prot$), the field jumps into a high-energy state, with local peaks spaced by roughly half the partial-wreath cycle period. 

\section{Polar Caps of Magnetism}\label{sec:polarcaps}
Also visible in Figure \ref{fig:tl_bphi_mid}(\textit{a}) is a weak azimuthal magnetic field at high latitudes on the order of tens of Gauss (see, for example, positive $\av{B_\phi}$ in the north and negative $\av{B_\phi}$ in the south during the interval from 3500 to 5000 $\prot$). These polar caps of magnetism fluctuate in amplitude and occasionally reverse polarity, but not with any regularity and with no discernible relationship to the fourfold-wreath or partial-wreath cycles. 

\begin{figure}
	\includegraphics{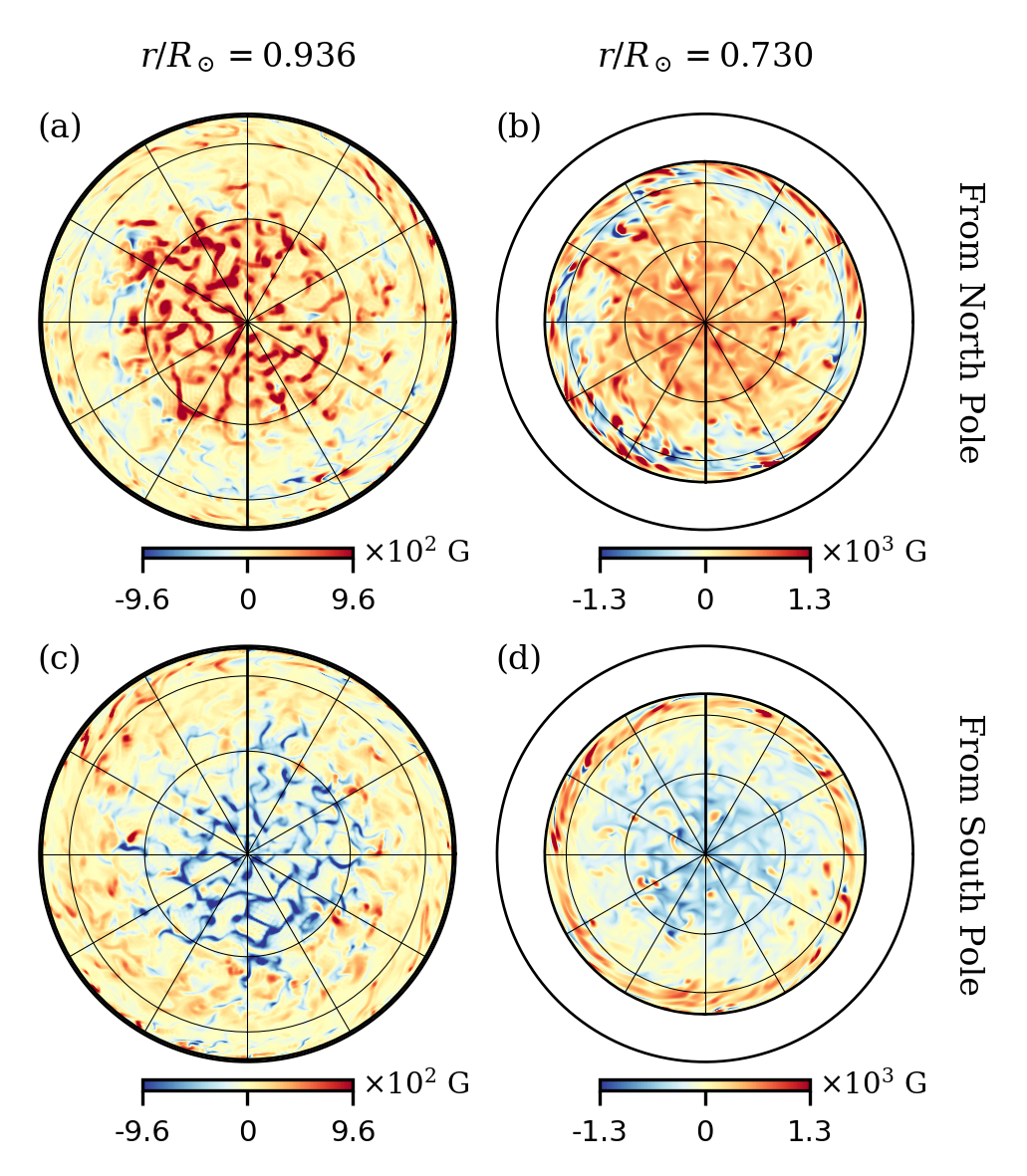}
	\caption{Spherical cuts of the instantaneous radial magnetic field $B_r$, viewed from the north pole (top row) and the south pole (bottom row). The instant shown is $t\approx4200\ \prot$, just after the first transition from partial-wreath to fourfold-wreath cycling when there is intense meridional field at the poles. Two depths are sampled: one near the top of the domain and one near the bottom.}
	\label{fig:ortho_polar}
\end{figure}

There is a significantly stronger polar cap associated with the meridional magnetic field $\bm{B}_{\rm{m}}$, with magnitude $\sim$1 kG. Figure \ref{fig:ortho_polar} shows the radial magnetic field when there are intense polar caps of magnetism. Near the top of the domain (left-hand column), the local structure of $B_r$ inherits the honeycomb network of downflows associated with the Taylor columns, while deeper down (right-hand column), the field is more homogeneous. At both depths shown, however, the complicated polar-cap structure on average resembles a dipolar field, with magnetic north lining up with the geometric north. The instantaneous views of $B_\theta$ (not shown) similarly display a dipolar structure, with $B_\theta$ positive near the poles in both hemispheres. 

\begin{figure*}
	\includegraphics{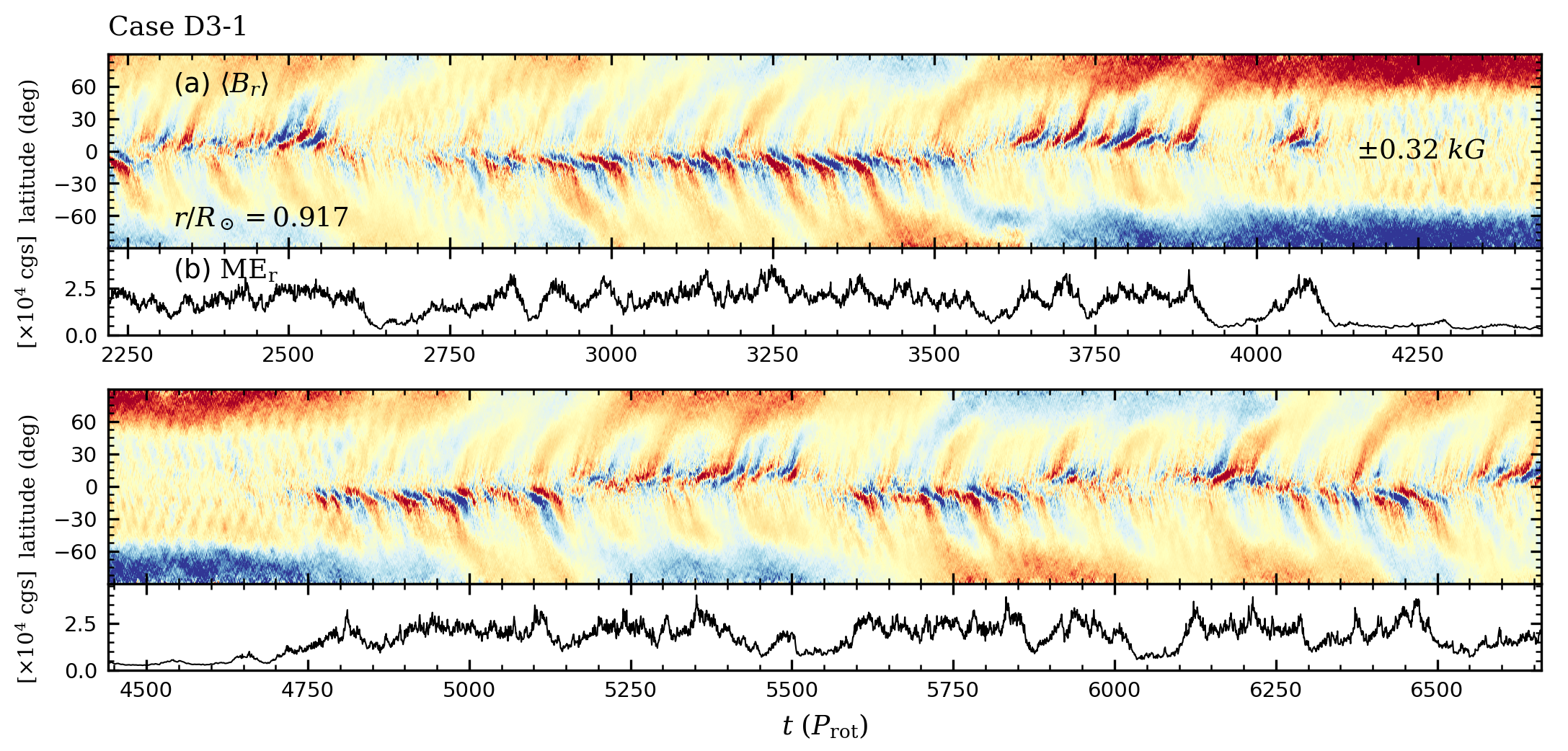}
	\caption{Time-latitude diagram of radial magnetic field $\av{B_r}$ for case D3-1 at $r/R_\odot = 0.916$ (upper CZ), and associated energy density at the same depth. The time interval sampled corresponds to the second and third panels of the extended time trace in Figure \ref{fig:tl_bphi_mid}.}
	\label{fig:tl_br_top}
\end{figure*}

Figure \ref{fig:tl_br_top} provides the time-latitude diagram for $\av{B_r}$ near the top of the domain during an interval in the middle of the simulation. There is clearly a strong dipolar component to the field at the poles. After a fairly quiescent initial interval, the dipolar field strengthens (at $t \approx 3750\ \prot$) a few hundred rotations before the fourfold-wreath cycle becomes dominant for the second time. The partial wreaths produce flux that is transported to the poles, seen in Figure \ref{fig:tl_br_top} as intermittent large plumes of red and blue. 

Not all the partial-wreath cycles produce field that propagates all the way to the poles, and there does not seem to be a one-to-one correspondence between polarity reversals of the partial wreaths and reversals in the polar caps. For nearly every partial-wreath cycle, however, the radial field propagates much further poleward than the $\pm20^\circ$ latitude limits that are maintained fairly consistently for $\av{B_\phi}$. 

The poleward migration of $\av{B_r}$ seen in Figure \ref{fig:tl_br_top} bears a striking resemblance to the concept the meridional circulation acting as a ``conveyor belt" of magnetism in the flux-transport (F-T) dynamo framework (e.g., \citealt{Charbonneau14}). In that picture, the meridional circulation advects the meridional component of magnetic field to the poles near the outer surface, where it accumulates and is eventually pumped back down to the equator in the deep CZ, also by the meridional circulation. This poloidal field is then sheared into toroidal field by the differential rotation, starting the cycle anew. In the F-T framework, the meridional circulation time thus sets the cycle period. 

However, case D3-1 is more complicated than the simple F-T model for two reasons. Firstly, the meridional circulation time for case D3-1 is on the order of $400\ \prot$, which is substantially longer than either the partial-wreath or fourfold-wreath cycle periods. Secondly, although the poles occasionally flip polarity, there is no clean correspondence between the polarity of the field at the poles and the polarity of the field in the equatorial regions during subsequent cycles.

\section{Evolution of  the Fourfold-Wreath Cycle}\label{sec:fourfold}
\begin{figure}
	\includegraphics{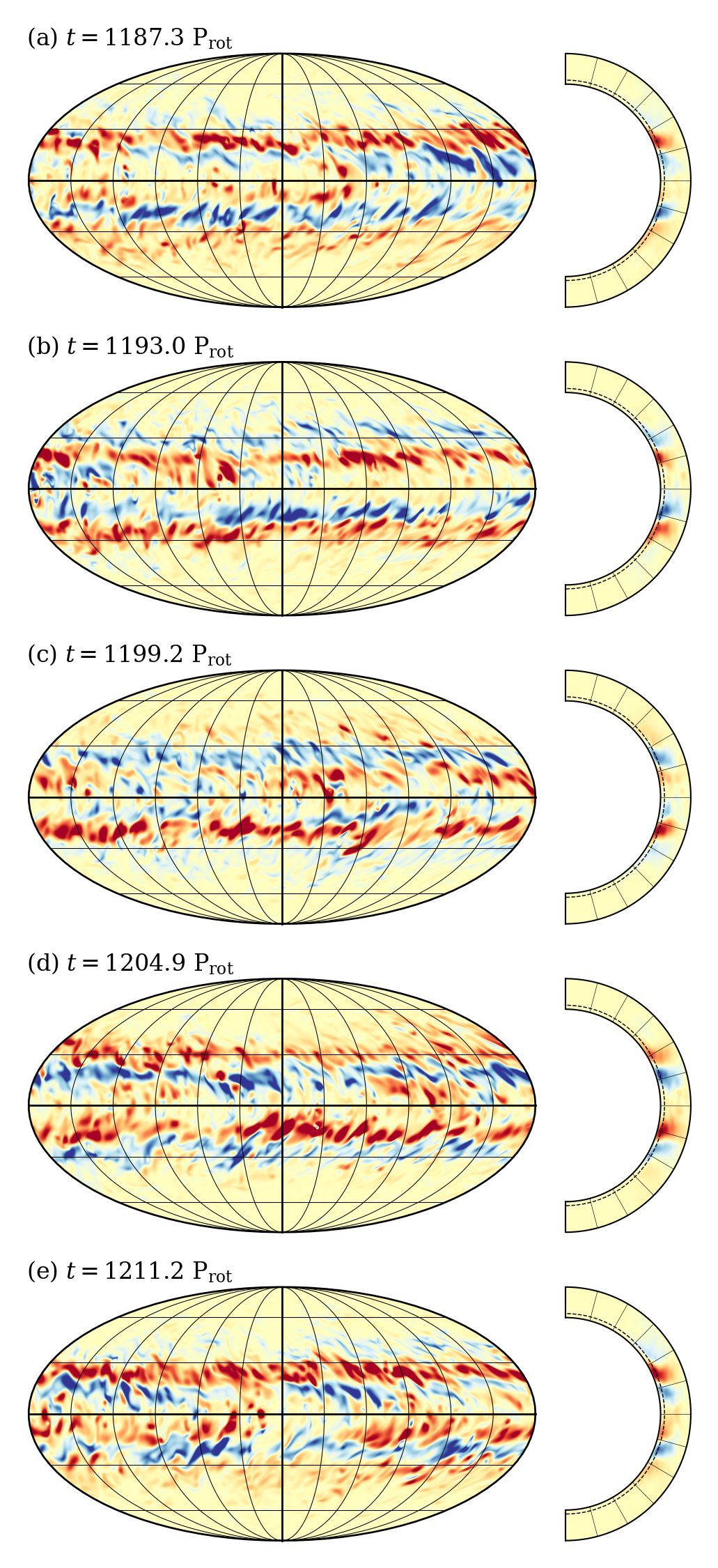}
	\caption{Mollweide snapshots (at $r/R_\odot=0.748$) of $B_\phi$ and meridional snapshots of $\av{B_\phi}$ for one fourfold-wreath cycle. Snapshots are separated by about one-quarter of a cycle ($\sim$6 $\prot$). The color map is identical to that of Figure \ref{fig:bphi_twodepths}(\textit{a}), with the Mollweide plots saturated at $\pm4.5$ kG and the longitude-average plots saturated at $\pm2.5$ kG.}
	\label{fig:bp_during_sunspot_cycle}
\end{figure}

Figure \ref{fig:bp_during_sunspot_cycle} shows the evolution of the azimuthal magnetic field $B_\phi$ during one reversal in the fourfold-wreath cycle. At any given time, each wreath has a complicated structure due to stretching and pummeling by the Taylor columns, but generally has the same sense for $B_\phi$ all the way around the sphere, indicating field lines that connect in a large torus. The accompanying snapshots of $\av{B_\phi}$ in the meridional plane indicate that the wreaths are strongest deep in the shell and are confined in radius to the lower half of the CZ and in latitude to within $\sim$$40^\circ$ of the equator. 

Each wreath marches steadily equatorward as the cycle progresses. As a wreath of one polarity reaches the equator, a new wreath of that same polarity appears at a higher latitude. In Figure \ref{fig:bp_during_sunspot_cycle}(\textit{a}), for example, there are two opposite-polarity wreaths in the northern hemisphere below $\sim$$30^\circ$-latitude (roughly equal in amplitude) and a negative-$\av{B_\phi}$ wreath that is just beginning to form above $\sim$$30^\circ$-latitude. By the time a given wreath migrates from mid-latitudes to within $\sim$$10^\circ$ of the equator, a new wreath of that same polarity has formed at mid-latitudes, completing the cycle. Therefore the configuration in Figure \ref{fig:bp_during_sunspot_cycle}(\textit{e}) has essentially returned to that of Figure \ref{fig:bp_during_sunspot_cycle}(\textit{a}). 

\begin{figure}
	\includegraphics{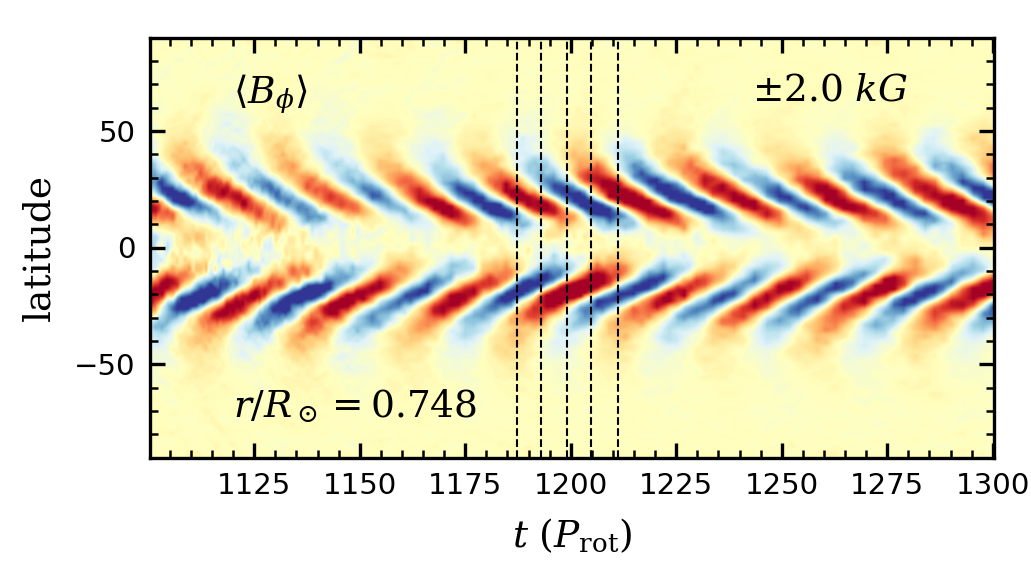}
	\caption{Close-up view of the time-latitude behavior of $\av{B_\phi}$ near the bottom of the CZ ($r/R_\odot=0.748$) during the interval marked by the vertical dashed lines in Figure \ref{fig:tl_bphi_mid}(\textit{a}). This interval (between 1100 and 1300 $\prot$), encompasses about eight fourfold-wreath cycles. The dashed lines indicate the times sampled by the series of Mollweide projections in Figure \ref{fig:bp_during_sunspot_cycle}.}
	\label{fig:zoom_tl_bphi_bot}
\end{figure}

Figure \ref{fig:zoom_tl_bphi_bot} shows a close-up view (in time-latitude space) of a small time interval containing about eight fourfold-wreath cycles (marked by the first set of dashed lines in Figure \ref{fig:tl_bphi_mid}(\textit{a})), including the instants sampled in Figure \ref{fig:bp_during_sunspot_cycle}.  Here the field is first seen to appear (at its weakest) at roughly $40^\circ$-latitude north and south and migrates steadily equatorward, attaining its maximum strength between $20^\circ$--$30^\circ$ north and south, while stopping short of the equator at about $10^\circ$ north and south. 

Interestingly, the southern hemisphere is often out of phase with the northern hemisphere and it is hard to pinpoint whether the fourfold-wreath structure is symmetric about the equator or antisymmetric. Indeed, the simulation exhibits both symmetric and antisymmetric tendencies at different times (for example, compare the symmetric configuration in Figure \ref{fig:bphi_twodepths}(\textit{a}) to the antisymmetric configurations shown in Figure \ref{fig:bp_during_sunspot_cycle}). At $t=1200\ \prot$ (horizontal center of Figure \ref{fig:zoom_tl_bphi_bot}), the wreath structure is clearly antisymmetric in the large.

\begin{figure}
	\includegraphics{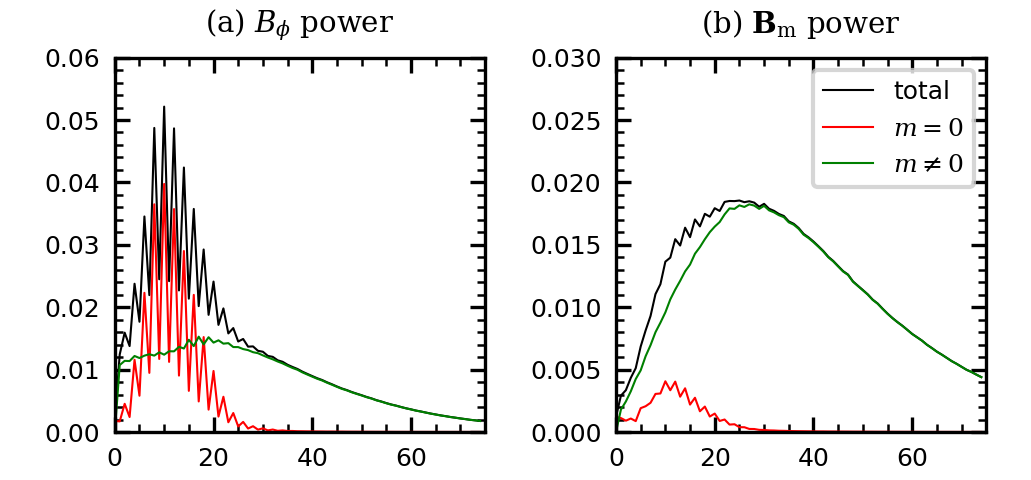}
	\caption{Power spectra of the magnetic field at $r/R_\odot=0.748$ during fourfold-wreath cycling, averaged between $t=0$ and $t=2200$ $\prot$. Separate curves indicate contributions from all $m$, $m=0$, and $|m|>0$. The sum of the total power over all $\ell$ equals 1. (\textit{a}) Spectrum of azimuthal magnetic field $B_\phi$. (\textit{b}) Spectrum of meridional magnetic field $\bm{B}_{\rm{m}} \define B_r\e_r+ B_\theta\e_\theta$.}
	\label{fig:pspec_sunspot}
\end{figure}

The symmetry of dynamo states in mean-field dynamo theory is usually quantified by the prevalence of the dipole ($\ell=1$) and quadrupole ($\ell=2$). Our dynamo cases, by contrast, are dominated by higher-order modes that serve to localize each wreath in space, leaving the dipolar and quadrupolar contributions extremely weak. Figure \ref{fig:pspec_sunspot} shows the magnetic-field power spectra, averaged over the initial interval during which the fourfold-wreath cycle dominates. The power in longitude-averaged magnetic fields (the $m=0$ power) peaks around $\ell=10$.  The power in the fluctuating magnetic fields ($|m|>0$) peaks at the smaller scales, around $\ell=20$ for $B_\phi^\prime$ and $\ell=25$ for $\bm{B}_{\rm{m}}^\prime$.  

The peak around $\ell=10$ is consistent with the fourfold-wreath structure as seen in Figures \ref{fig:bphi_twodepths}(\textit{a}) and \ref{fig:bp_during_sunspot_cycle}: each wreath in a given hemisphere has a latitudinal extent of about $15^\circ$, or roughly one-tenth of the $180^\circ$ in latitude covered by the full sphere. We note that for the longitude-averaged fields, the dipolar and quadrupolar modes ($\ell=1$ and $2$, respectively) are weaker by a factor of $\sim$$10$ compared to any of the modes near $\ell=10$.

\begin{figure*}
	\includegraphics{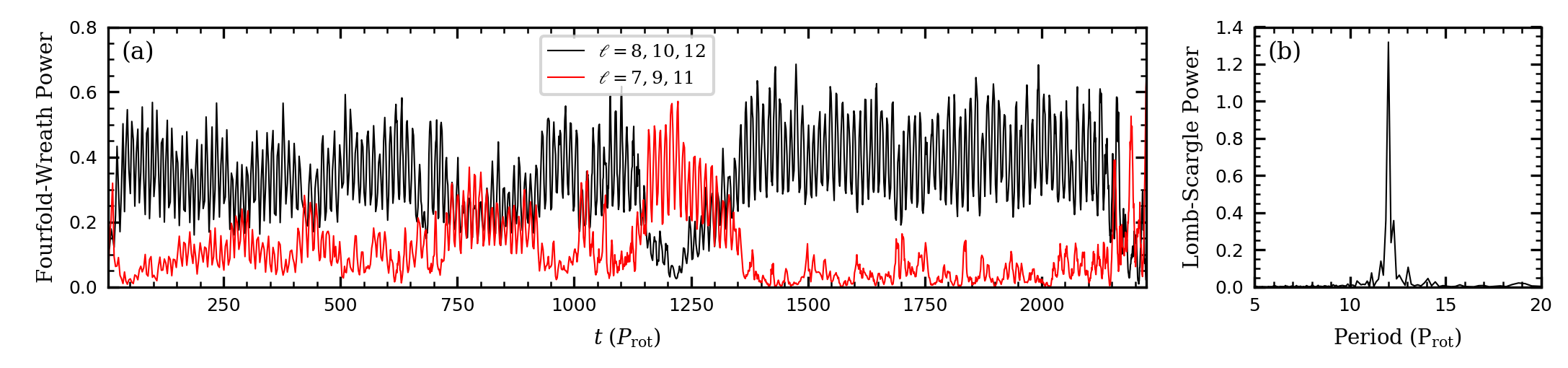}
	\caption{(\textit{a}) Temporal evolution of the power for $\av{B_\phi}$ at $r/R_\odot=0.748$ in the modes $\ell=8,10,12$ (black curve) and $\ell=7,9,11$ (red curve). These are the even/odd modes near the peak of the spectrum at $\ell\approx10$, which contain most of the power. Power is shown during the initial interval for which the fourfold-wreath cycle is dominant, coincident with the top panel in Figure \ref{fig:tl_bphi_mid}. (\textit{b}) Lomb-Scargle periodogram of the $\ell=7$--$12$ contributions to $\av{B_\phi}$ during the interval shown in (\textit{a}). Normalized Lomb-Scargle power in each frequency component is shown versus period, with a clear peak occurring at $P_{\rm{energy}} \equiv 12.0\ \prot$.}
	\label{fig:pspec_vs_time_sunspot}
\end{figure*}

In Figure \ref{fig:pspec_vs_time_sunspot}(\textit{a}), we show the temporal behavior of the power in the longitude-averaged ($m=0$) field $\av{B_\phi}$ around $\ell=10$ deep in the shell during the fourfold-wreath cycle. Most of the time the fourfold-wreath structure is symmetric about the equator, with the even modes $\ell=8,10,12$ containing up to 60\% of the total power for $\av{B_\phi}$. The two hemispheres are out of phase, however, and the symmetric (odd) and antisymmetric (even) modes can sometimes be roughly equal (e.g., near $t=800\ \prot$). Near $t=1200\ \prot$, the antisymmetric modes dominate, as we noted in discussing Figure \ref{fig:zoom_tl_bphi_bot}. 

The $m=0$ power around $\ell=10$ is is a proxy for the total energy contained in the fourfold-wreath structure. The oscillations in Figure \ref{fig:pspec_vs_time_sunspot}(\textit{a}) thus show that the fourfold-wreath energy oscillates periodically. In Figure \ref{fig:pspec_vs_time_sunspot}(\textit{b}),  we decompose the time series for all the power contained around $\ell=10$, both symmetric (even) and antisymmetric (odd) modes, into its frequency (or period) components. There is clearly one period that is dominant, which corresponds to an energy cycle of $\rm{P_{energy}} = 12.0\ \prot$. Since the polarity alternates between energy cycles, the fourfold-wreath cycle period is
\begin{align}
\rm{P_{fourfold}} =2\ \rm{P_{energy}} = 24.0\ \prot, 
\label{eq:actual_sunspot_period}
\end{align}
similar to the estimate of $\sim$$25\ \prot$ mentioned previously. 

\section{Evolution of  the partial-wreath cycle}\label{sec:partial}
 The partial-wreath component (when present) is strongest at mid-depth.  From Figure \ref{fig:tl_bphi_mid}(\textit{a}), the partial-wreath structure first appears around $t=2100\ \prot$ and cycles between positive-$\av{B_\phi}$-dominant and negative-$\av{B_\phi}$-dominant partial-wreath pairs. Unlike the fourfold-wreath structure (which continues cycling regularly even when the partial-wreath structure dominates), the partial-wreath cycling is intermittent, turning off altogether at times, but always returning.  There is some variation in the time-latitude behavior of $\av{B_\phi}$ from cycle to cycle, but in general, $\av{B_\phi}$ stays confined to within $\sim$$20^\circ$ of the equator in a given hemisphere and shows a tendency to propagate toward higher latitudes with time.
 
Even when the partial-wreath cycling is dominant, the equatorward-propagating fourfold-wreath cycle still persists with its regular period of 24.0 $\prot$ unchanged. However, the fourfold-wreath structure is of much weaker amplitude and is less coherent in the hemisphere where the partial wreaths reside. There is thus a wide dynamical range in the amplitudes of the magnetic field overall. The partial-wreath cycle is associated with longitude-averaged field strengths of $\sim$10 kG, and when the partial-wreath cycle is dominant, the fourfold-wreath cycle's longitude-averaged field strengths are reduced to several hundred G from $\sim$2 kG. 

\begin{figure}
	\includegraphics{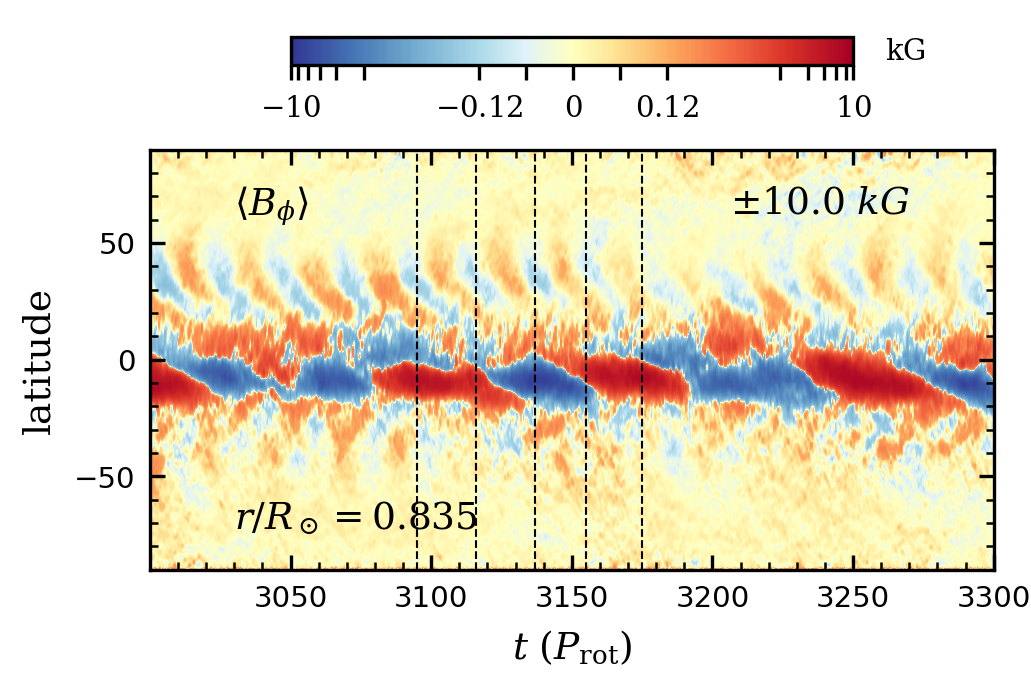}
	\caption{Close-up view of the time-latitude behavior of $\av{B_\phi}$ during the interval marked by the first set of vertical dashed lines in Figure \ref{fig:tl_bphi_mid}(\textit{a}). This interval (between 3000 and 3300 $\prot$), encompasses about four partial-wreath cycles. The vertical dashed lines denote the instances sampled by the volume renderings of $B_\phi$ in Figure \ref{fig:volrender}.}
	\label{fig:zoom_tl_bphi_mid}
\end{figure}

Figure \ref{fig:zoom_tl_bphi_mid} shows a close-up view of the partial-wreath mode when it is cycling in the south. Here the propagation of $\av{B_\phi}$ away from the equator is clear. For any given cycle, the partial-wreath structure begins roughly at the equator and migrates toward mid-latitudes, stopping at $\sim$$\pm20^\circ$. Displaying $\av{B_\phi}$ as in Figures \ref{fig:tl_bphi_mid}(\textit{a}) and \ref{fig:zoom_tl_bphi_mid} gives a sense of the complex temporal evolution, but it must be complemented by examining the spatial structures of $B_\phi$ before they are averaged. This is particularly true for the partial-wreath cycling, where what appear as polarity reversals in $\av{B_\phi}$ are actually modulations in the the relative strengths of the partial wreaths.

\begin{figure}
	\includegraphics{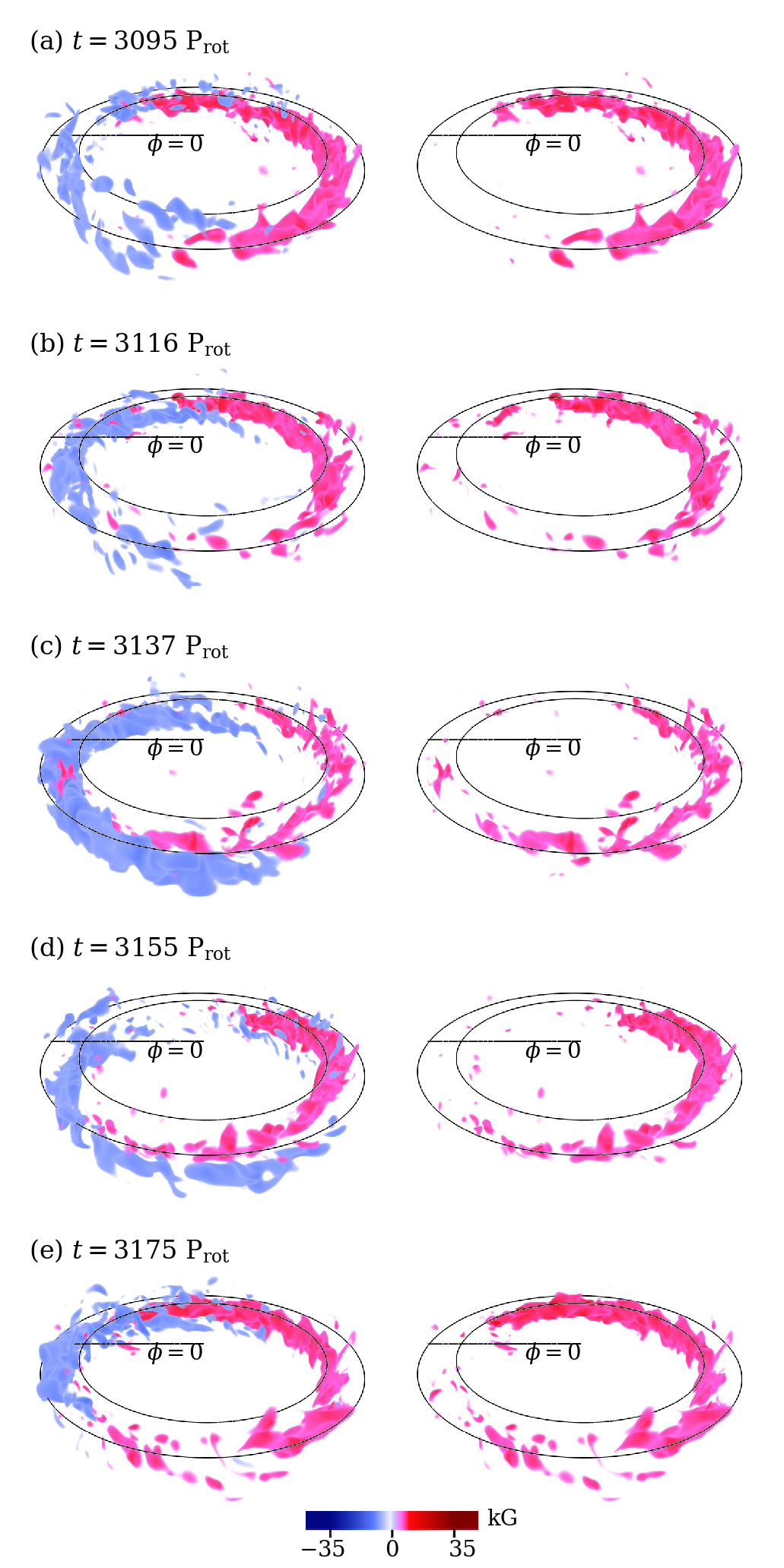}
	\caption{Volume rendering of $B_\phi$ throughout one partial-wreath cycle, tracked in a frame rotating at 32 nHz faster than the frame rotation rate $\Omega_0$, approximately equal to the equatorial rotation rate at mid-depth.  Only strong fields ($|B_\phi| >$  5 kG) are depicted. To show the field everywhere in 3D space and simultaneously emphasize the strong fields, transparency varies linearly with field strength, with structures for which $|B_\phi|\geq$ 35 kG being completely opaque. The first row (panel \textit{a}) samples the field when the red partial wreath is dominant, and subsequent samples (panels \textit{b}--\textit{e}) are separated by roughly one-quarter of the partial-wreath cycle period. The left-hand column shows the full $B_\phi$ profile and the right-hand column shows only positive $B_\phi$. The view is from $\sim$$30^\circ$ north of the equator and $90^\circ$ longitude. The spherical-shell boundaries in the equatorial plane and the location of 0$^\circ$-longitude are marked by the black curves.}
	\label{fig:volrender}
\end{figure}

 To determine in detail how the partial-wreath structure modulates over the course of a cycle, we show in Figure \ref{fig:volrender} a series of volume renderings of the azimuthal field, in a tracking frame rotating somewhat faster than $\Omega_0$. In each snapshot, the positive-$B_\phi$ and negative-$B_\phi$ partial wreaths are centered roughly around $\phi = 210^\circ$ and $\phi = 30^\circ$ in longitude, respectively, although ``islands" of magnetism associated with each wreath extend around the whole sphere. In the first snapshot (panel \textit{a}), the positive-$B_\phi$ partial-wreath is dominant. Its amplitude and longitudinal extent then steadily decline over the next half-cycle while the negative-$B_\phi$ partial-wreath becomes dominant (panels \textit{b} and \textit{c}). In the second half of the cycle (panels \textit{d} and \textit{e}), the positive-$B_\phi$ wreath expands and strengthens until it is dominant again and the configuration is roughly the same it was at the beginning of the cycle. In the longitudinal average (e.g., Figure \ref{fig:tl_bphi_mid}(\textit{a})), much cancellation occurs, but the sign of $\av{B_\phi}$ still reflects which partial wreath is dominant. 

\begin{figure}
	\includegraphics{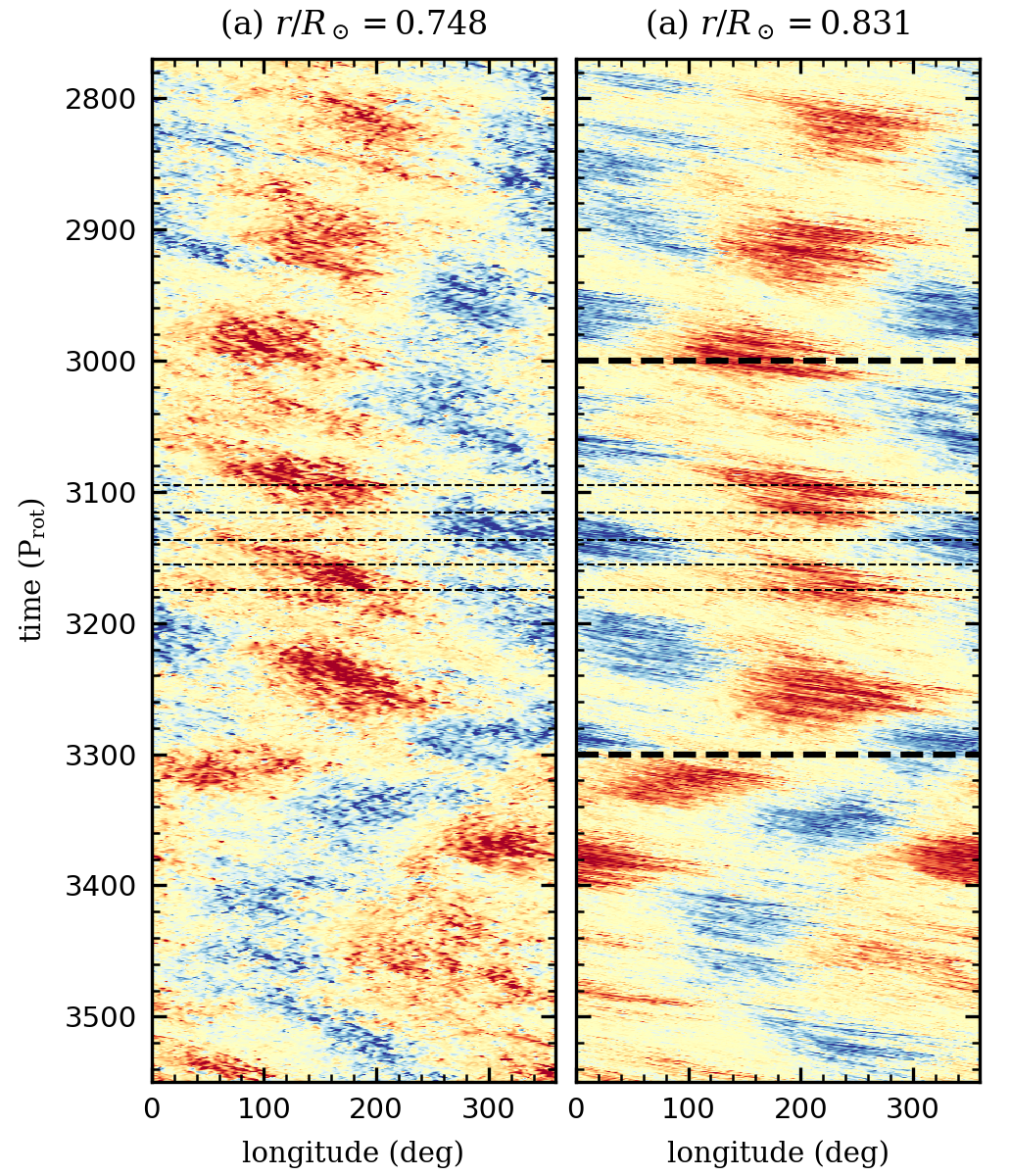}
	\caption{Time-longitude trace of $B_\phi$ for $10^\circ$-south latitude (\textit{a}) deep in the shell and (\textit{b}) at mid-depth,tracked at the same rotation rate as in Figure \ref{fig:volrender}.  The color map saturation values are $\pm12$ kG in (\textit{a}) and $\pm25$ kG in (\textit{b}). The interval shown in close-up view in Figure \ref{fig:zoom_tl_bphi_mid} is marked by the thick horizontal dashed lines in panel (\textit{b}). The instances sampled in Figure \ref{fig:volrender} are marked by the thinner dashed lines in both panels.}
	\label{fig:tlon}
\end{figure}

In order to track the partial-wreath structure over multiple cycles, we trace $B_\phi$ with respect to time and longitude in a frame rotating at the same rate as in Figure \ref{fig:volrender}. Figure \ref{fig:tlon} shows the time-longitude trace for $B_\phi$ at $10^\circ$-south latitude and at two depths, during an interval when the partial-wreath structure is stronger than the fourfold-wreath structure and cycling in the south. The choice of $10^\circ$-south latitude cuts the partial wreaths roughly in their core, where the field strengths are highest.

Approximately the same propagation pattern in time-longitude space occurs both at mid-depth and deep in the shell, despite the field being tracked at the same rate for both depths. This rate is close to the equatorial fluid rotation rate at mid-depth, suggesting that the whole asymmetric wreath structure is ``anchored" in the middle of the shell and moves through the fluid as a single entity. At mid-depth, the field pattern is uniformly located at a higher longitude than the field pattern deep down. This displacement does not change with time, suggesting that it is due to the inherent geometry of the wreath rather than the shearing action of the flow.  

In the rotating frame chosen, the field structure is approximately stationary. Each partial wreath (either positive-$B_\phi$ or negative-$B_\phi$) is modulated in amplitude over a period of $\sim$$80\ \prot$ (the same as the cycle period identified in the time-latitude trace of $\av{B_\phi}$), but remains basically at a fixed longitude. Furthermore, the positive-$B_\phi$ and negative-$B_\phi$ wreaths are out of phase with another. At any given time, there are thus two partial wreaths, but in general when the positive-$B_\phi$ wreath is strong, the negative-$B_\phi$ wreath is weak, and vice versa.

The partial-wreath structure is still not exactly stationary even when tracked in the chosen frame. Figure \ref{fig:tlon} shows a weak zigzag pattern for each the positive-$B_\phi$ and negative-$B_\phi$ partial wreaths, implying that the propagation rate of the wreath through the fluid is not constant with time, or that longitudinal extents of each partial wreath shrink and expand throughout the cycle.

In light of Figures \ref{fig:volrender} and \ref{fig:tlon}, the ``polarity reversals" in $\av{B_\phi}$ during the partial-wreath cycle are in fact accomplished by each partial wreath weakening and strengthening mostly \textit{in place}, with the positive-$B_\phi$ partial wreath oscillating in time 180$^\circ$ out of phase with negative one. This stands in contrast to the reversals of the fourfold-wreath cycles, which are achieved by full wreaths migrating equatorward from mid-latitudes and getting replaced by new wreaths of the opposite polarity. 

\subsection{Partial-wreath cycle period and hemispheric asymmetry}
The partial-wreath cycle period is roughly $80\ \prot$, as identified previously, but it is substantially irregular. We determine the period more formally by calculating the periodicities associated with the relative energy content in the two partial wreaths:
\begin{align}
A_{\rm{partial}} \equiv \frac{B_\phi^+ - B_\phi^-} {B_\phi^+ + B_\phi^-},
\label{eq:apartial}
\end{align}
the ``+" and ``$-$" superscripts referring to instantaneous rms averages of the field strength over the regions where $B_\phi$ is positive and negative (respectively), thus measuring the energy content in the positive and negative partial wreaths separately. The mean in the rms is taken over longitude, low latitudes between $\pm25^\circ$-latitude, and all depths. Unlike the linear longitudinal average $\av{B_\phi}$, the rms-average $A_{\rm{partial}}$ is more sensitive to the strongest fields, and thus better determines the relative amplitudes of the partial wreaths as visualized in Figure \ref{fig:volrender}. 

\begin{figure*}
	\includegraphics{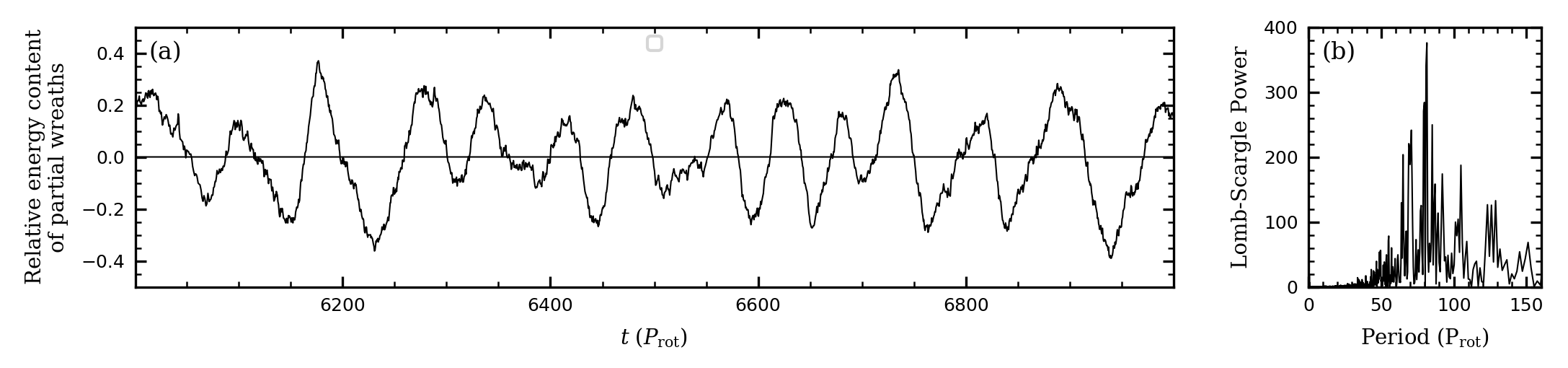}
	\caption{(\textit{a}) Temporal evolution of $A_{\rm{partial}}$ (Equation \eqref{eq:apartial}) during the interval between $6000\ \prot$ and  $7000\ \prot$ in case D3-1. (\textit{b}) Normalized Lomb-Scargle periodogram of $A_{\rm{partial}}$ for case D3-1 during the interval from $5100\ \prot$ until the end of the simulation, during which time the partial-wreath pair is consistently stronger than the fourfold-wreath structure.}
	\label{fig:partial_wreath_time_trace}
\end{figure*}

Figure \ref{fig:partial_wreath_time_trace}(\textit{a}) shows the relative energy content over an interval for which no period is easily identifiable in the time-latitude trace of $\av{B_\phi}$ in Figure \ref{fig:tl_bphi_mid}(\textit{a}), but the partial-wreath cycle is still visible in the well-defined peaks and troughs of $A_{\rm{partial}}$. In Figure \ref{fig:partial_wreath_time_trace}(\textit{b}), we show the periodogram associated with $A_{\rm{partial}}$. There is a distinctive peak in power at
\begin{align}
\rm{P_{partial}} \equiv 81.5\ \prot, 
\label{eq:partial_wreath_period}
\end{align}
which we define to be the (dominant) partial-wreath cycle period. Due to the irregularity of the partial-wreath cycle---caused by some cycles being longer or shorter than $\rm{P_{partial}}$, as well as significant amplitude and phase modulation---there is a substantial spread in the periods for which the Lomb-Scargle power is large. The dominant period identified in Equation \eqref{eq:partial_wreath_period} is thus consistent with the previous estimate $\sim$80 $\prot$ from a visual examination of Figure \ref{fig:tl_bphi_mid}(\textit{a}).

The partial wreaths tend to reside predominantly in one hemisphere or the other, although they are not strictly confined and can in some instances cross the equator. Nonetheless, this phenomenon leads to a substantial \textit{hemispheric asymmetry}, which we define for our dynamo models as
\begin{align}
A_{\rm{hem}} \equiv \frac{B_\phi^{\rm{N}} - B_\phi^{\rm{S}} }{B_\phi^{\rm{N}}  + B_\phi^{\rm{S}} },
\label{eq:ahem}
\end{align}
the ``N" and ``S" superscripts referring to instantaneous rms averages of the field strength over the northern and southern hemispheres (respectively). The mean in the rms is taken over longitude, low latitudes (between the equator and $25^\circ$-latitude for the northern hemisphere and comparably for the southern hemisphere), and all depths. 

\begin{figure*}
	\includegraphics{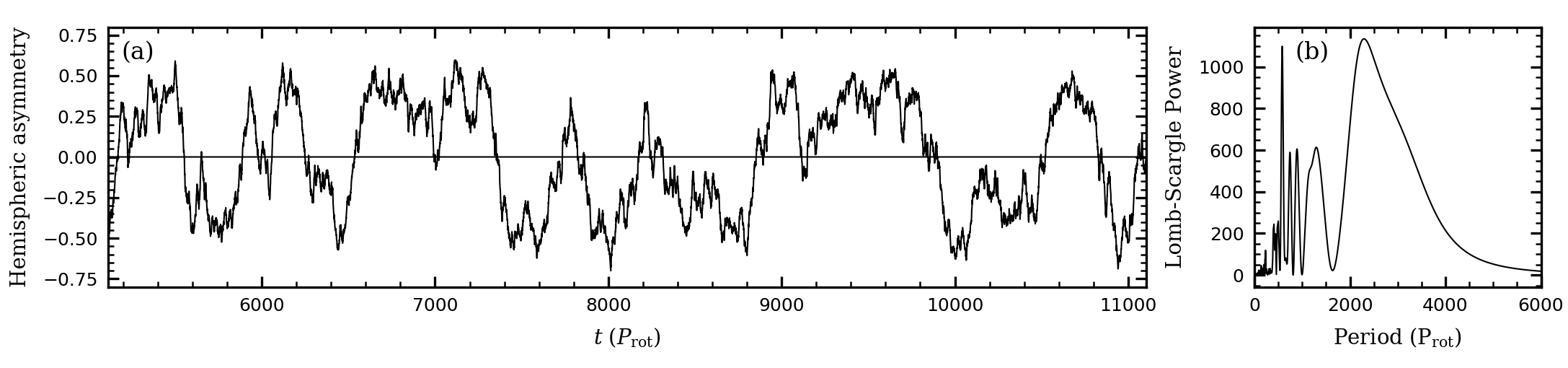}
	\caption{(\textit{a}) Temporal evolution of $A_{\rm{hem}}$ (Equation \eqref{eq:ahem}) during the interval between $5100\ \prot$ and  the end of the simulation in case D3-1.  (\textit{b}) Normalized Lomb-Scargle periodogram of $A_{\rm{hem}}$ for case D3-1 during the same interval as in panel (\textit{a}).}
	\label{fig:hemispheric_asymmetry}
\end{figure*}

Figure \ref{fig:hemispheric_asymmetry}(\textit{a}) shows the temporal behavior of the hemispheric asymmetry for case D3-1 during roughly the latter half of the simulation, when the partial-wreath cycle is usually substantially stronger than the fourfold-wreath cycle. The dominant hemisphere switches between north and south chaotically, with no single period easily seen. We have confirmed this aperiodic nature by computing the periodogram associated with $A_{\rm{hem}}$ shown in Figure \ref{fig:hemispheric_asymmetry}(\textit{b}): there are a wide range of periods that are relevant, with no obvious peak in the power spectrum. What is clear from the broad peak near $2000\ \prot$ in the periodogram (and from a visual inspection of Figure \ref{fig:hemispheric_asymmetry}(\textit{a})) is that asymmetry of a single sign---corresponding to partial wreaths cycling in preferentially in a single hemisphere---can persist for long time scales, up to $\sim$1000 $\prot$.

\section{Bistability Trends with Higher Magnetic Prandtl Number}\label{sec:higher_pm}
\begin{figure}
	\includegraphics{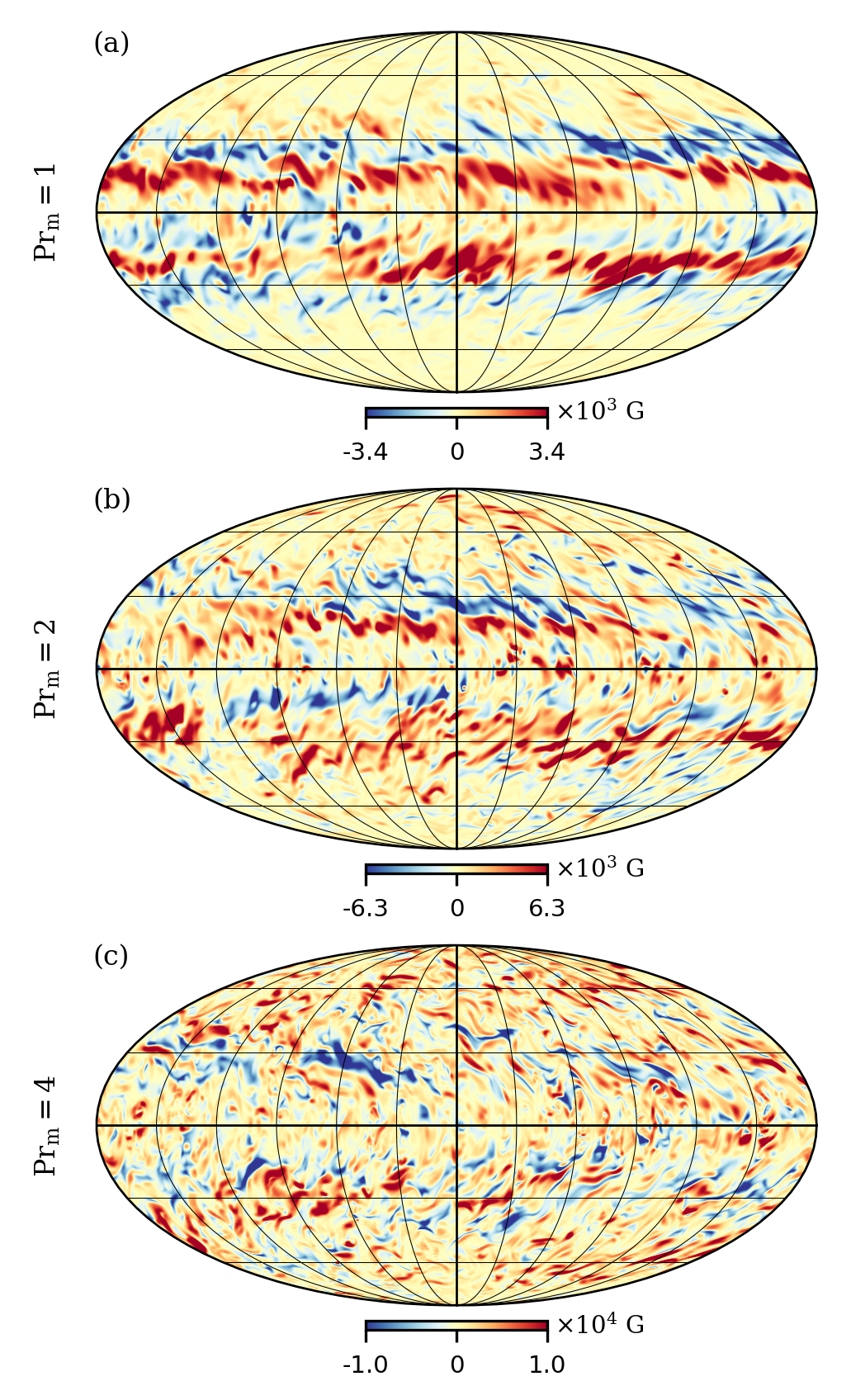}
	\caption{Snapshots on spherical surfaces of the azimuthal magnetic field $B_\phi$ shown in global Mollweide projection for the three dynamo cases D3-1, D3-2, and D3-4 with increasing $\prm$ from (\textit{a}) to (\textit{c}). Each model is sampled near the bottom of the shell ($r/R_\odot = 0.748$) at an early time within the dynamo simulation (less than 100 $\prot$ after appreciable magnetism develops).}
	\label{fig:moll_vs_pm}
\end{figure}

Because we keep the viscous diffusivity fixed while lowering the magnetic diffusivity, the magnetic structures grow increasingly more complex the higher the magnetic Prandtl number.  Figure \ref{fig:moll_vs_pm} contains snapshots of the azimuthal magnetic field $B_\phi$ (in Mollweide projection) near the beginning of each of the three dynamo cases. At these early times, the structure of $B_\phi$ consists largely of the fourfold wreaths. As $\prm$ is increased, the magnetic structures become more shredded and dominated by small-scale features. The local field strength increases, but the global coherence of the field decreases.

\begin{figure*}
	\includegraphics{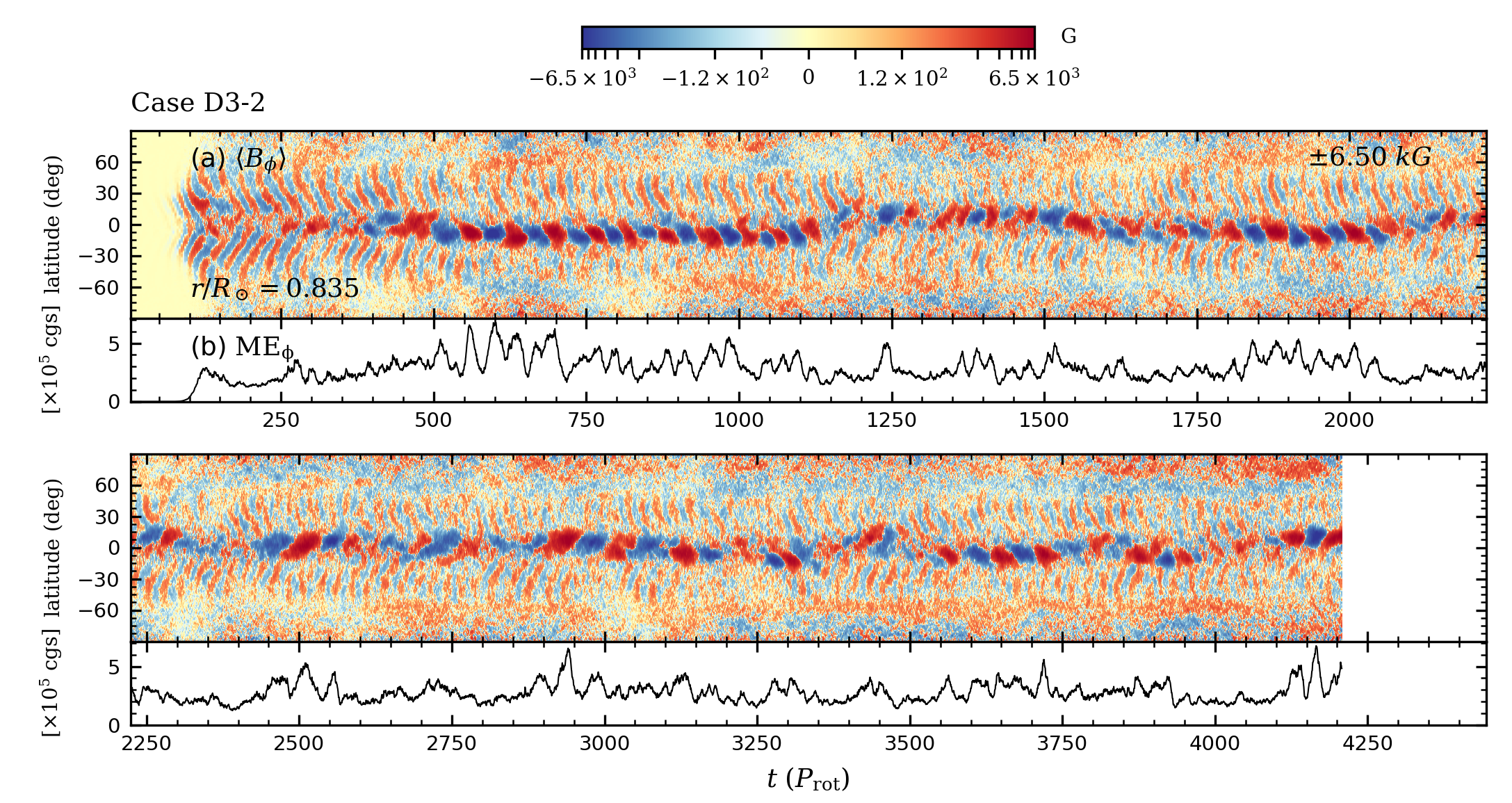}
	\caption{Extended time-latitude diagram of azimuthal magnetic field $\av{B_\phi}$ and energy trace for $\rm{ME_\phi}$ at mid-depth ($r/R_\odot=0.835$) for case D3-2 (at $\prm=2$), similar to Figure \ref{fig:tl_bphi_mid}. Here the the fourfold-wreath and partial-wreath cycles are seen to coexist for most of the simulation, with the partial-wreath cycle in particular being less intermittent than in case D3-1.}
	\label{fig:tl_bphi_mid2}
\end{figure*}

Although bistability (in the form of the fourfold and partial-wreath cycles) is most clearly evident in case D3-1, similar behavior can be seen in the higher-$\prm$ cases D3-2 and D3-4. In Figure \ref{fig:tl_bphi_mid2}, we show extended time-latitude diagrams for case D3-2, with the time axes and color map scaled to be directly comparable to Figure \ref{fig:tl_bphi_mid}. Case D3-2 exhibits bistability with largely the same characteristics as case D3-1, namely a fourfold-wreath cycle with regular polarity reversals and a uni-hemispherical partial-wreath cycle with irregular polarity reversals. The fourfold-wreath cycle has a weaker signal, however, due to the finer-scale magnetic structures. The system destabilizes after only $\sim$250 $\prot$ to launch the partial-wreath cycle, which is less intermittent than in case D3-1, wandering between the northern and southern hemispheres without ever fully shutting down. 

From Figure \ref{fig:moll_vs_pm}(\textit{c}), the fourfold-wreath structure in case D3-4 is less striking than for the lower-magnetic-Prandtl number cases. Each wreath is also significantly tilted in latitude and thus there is very little signal left after a longitudinal average. The partial wreaths, which appear later in the simulation, are also less coherent. Nonetheless, the fourfold-wreath and partial-wreath cycles in case D3-4 can still be detected in sequences of Mollweide projections. 

\begin{table*}
	\caption{Diagnostic parameters, grid resolution, and simulation run times for all cases.}\label{tab:nond}
	\centering
	\begin{tabular}{l l l l l l   l l l l l l }
		\hline\hline
		Case  & $\prm$ & $\eta(\ro)$  & Ro & Re &  $\rem$ & $\Delta\Omega/\Omega_0$ & $(N_r, N_\theta, N_\phi)$ & $\ell_{\rm{max}}$ & Diffusion time & Run time & Run time \\
		\hline
		H3  & - & - & 0.0599 & 56.4 & - & 0.228 & (96, 384, 768) & 255 & 115 $\prot$  &  6390 $\prot$ & 55.6 TDT \\
		D3-1  & 1 & $\sn{3.00}{12}$ & 0.0609 & 58.3 & 76.5 & 0.183 & (96, 384, 768) & 255 & 115 $\prot$  & 11100 $\prot$  & 96.7  MDT \\		
		D3-2    & 2  & $\sn{1.50}{12}$& 0.0609 & 57.0 & 101 & 0.127  & (96, 384, 768) & 255 & 230  $\prot$  & 4210 $\prot$ & 18.3  MDT\\		 	
		D3-4   & 4 & $\sn{0.75}{12}$  & 0.0599 & 53.0 & 131 & 0.063 &  (128, 576, 1152) & 383 & 460 $\prot$  &  2140 $\prot$  & 4.65  MDT\\	 		
		\hline
	\end{tabular}
	\tablecomments{See Appendix \ref{ap:nond} for the definition of the diffusion times and non-dimensional numbers. For case H3 we quote the thermal diffusion time (TDT) and for the dynamo cases we quote the magnetic diffusion time (MDT). The maximum spherical-harmonic degree used in the horizontal expansion is taken to be $\ell_{\rm{max}} = 2N_\theta/3 - 1$ for dealiasing.}
\end{table*}

\begin{table*}
	\caption{Volume-averaged energy budgets for cases H3, D3-1, D3-2, and D3-4. ``KE" and ``ME" refer to the kinetic and magnetic energy densities, respectively.}\label{tab:en}
	\centering
	\begin{tabular}{l l l l   l l l l   l}
		\hline\hline
		Case  & KE/$10^6$& $\rm{KE_\phi}$/$10^6$ & $\rm{KE_{\rm{m}}}$/$10^6$ & $\rm{KE_c}$/$10^6$  & ME/$10^5$& $\rm{ME_\phi}$/$10^5$ & $\rm{ME_m}$/$10^5$ & $\rm{ME_c}$/$10^5$ \\
		\hline
		H3 & 19.3 & 17.5 (91.1\%) & 0.0053 (0.027\%) & 1.70 (8.83\%) & - & - & - & -\\
		D3-1 & 13.5 & 11.7 (86.7\%) & 0.0053 (0.039\%) & 1.79 (13.2\%) & 2.15 & 0.161 (7.50\%) & 0.015 (0.69\%) & 1.98 (91.8\%)\\		
		D3-2 & 7.89 & 6.12 (77.6\%) & 0.0048 (0.061\%) & 1.76 (22.3\%) & 3.79 & 0.162 (4.26\%) & 0.029 (0.77\%) & 3.60 (95.0\%)\\		
		D3-4 & 3.52 & 1.88 (53.4\%) & 0.0038 (0.11\%) & 1.64 (46.5\%) & 6.15 & 0.051 (0.82\%) & 0.031 (0.50\%) & 6.07 (98.7\%)\\		
		\hline
	\end{tabular}
	\tablecomments{Units are c.g.s., with a common exponent divided out. The subscripts $\phi$, $\rm{m}$, and $\rm{c}$ refer the energy contributions from the mean azimuthal field components ($\av{v_\phi}$ and $\av{B_\phi}$), the mean meridional field components ($\av{\bm{v}_{\rm{m}}}$ and $\av{\bm{B}_{\rm{m}}}$), and the convective components ($\bm{v}^\prime$ and $\bm{B}^\prime$), respectively. The fraction of the total energy (no subscripts; magnetic and kinetic energies considered separately) from each contribution is shown as the percentage in parentheses.
	}
\end{table*}

Each of the higher-$\prm$ cases has substantial polar caps of magnetism, which occasionally reverse polarity but are not clearly tied to either of the two cycles. In time-latitude diagrams for $\av{B_r}$ and $\av{B_\theta}$ (not shown), there is a clear tendency for the meridional field to break off from the partial wreaths and move to the poles, just as in Figure \ref{fig:tl_br_top} for case D3-1. 

The biggest global effect of increasing the magnetic Prandtl number is to lower the overall differential rotation, while keeping the shape of the rotation profile in the meridional plane largely fixed. Table \ref{tab:nond} contains the values of some of the diagnostic non-dimensional parameters (as well as the grid resolution and various timescales) characterizing each system. We quantify the strength of the differential rotation as the difference at the outer surface of the rotation rate between the equator and 60$^\circ$-latitude, normalized by the frame rotation rate: $\Delta\Omega/\Omega_0 \equiv [\Omega(\ro, \pi/2) - \Omega(\ro, \pi/6)]/\Omega_0$. For comparison, for the solar rotation rate determined through helioseismology, $\Delta\Omega/\Omega_0 = 0.197$, if we take $\Omega_0=\Omega_\odot$ and $\ro$ to be the radius just below near-surface shear layer \citep{Howe00}. The magnitude of the differential rotation in H3 is quite strong at 0.228---even greater than the solar value. As $\prm$ increases, $\Delta\Omega/\Omega_0$ is steadily weakened by the presence of small-scale magnetic structures. These counteract the Reynolds stress produced by the Taylor columns (which tend to drive a strong differential rotation) with feedbacks from the non-axisymmetric Maxwell stress. 

From Table \ref{tab:nond}, modifying the magnetic Prandtl number has little effect on the resulting level of rotational constraint (parameterized by the Rossby number) or the level of turbulence (parameterized by the Reynolds number). This agrees with the fact that the hydrodynamic structures (in terms of their distribution of length-scales and amplitudes) in the dynamo cases are largely similar to those of the hydrodynamic case H3. The main effect of modifying $\prm$ comes from the degree of complexity in the magnetic structures, as seen by the steady increase of $\rem$ with $\prm$. 

The bulk energy budget for the simulation suite is shown in Table \ref{tab:en}. Case H3 has both the highest kinetic energy from the differential rotation ($\rm{KE_\phi}$) and the highest kinetic energy overall. For the dynamo cases, as the magnetic Prandtl number is increased, the energy in the meridional circulation ($\rm{KE_m}$) and convection ($\rm{KE_c}$) stays roughly the same as in case H3, while the energy in the differential rotation ($\rm{KE_\phi}$) sharply decreases. 

The magnetic energy density (which is dominated by the fluctuating contribution $\rm{ME_c}$ for all three dynamo cases) steadily increases with increasing $\prm$, at the expense of the energy in the differential rotation. The energy in the mean magnetic fields $\rm{ME_\phi}$ and $\rm{ME_m}$ seems to vary in no clear way with $\prm$, although the mean energy in the azimuthal field ($\rm{ME_\phi}$) is significantly less in case D3-4, consistent with the shredded profile of $B_\phi$ seen in Figure \ref{fig:moll_vs_pm}(\textit{c}).


\section {The Dynamical Origins of Each Cycling Mode}\label{sec:dyn}
The real 22-year sunspot cycle must be the result of a complex interaction between turbulent convection, rotation, and magnetism. Our simulations, although far less turbulent than the Sun, have the advantage that the dynamical origins of each of the two cycling modes identified can be explored in a fair amount of detail. This is because we can probe each term in the dynamical equations with high spatio-temporal resolution and accuracy, something we are unable to do for the Sun. We devote this section to describing the dynamics that lead to the partial-wreath and fourfold-wreath cycles in case D3-1. We focus the discussion around the dominant terms in the induction equation \eqref{eq:ind} and their physical origins. 

We note at the outset that for both cycles, there is very little temporal modulation in either the differential rotation (DR; the main driver of the dynamo through the $\Omega$-effect) or the meridional circulation (MC). This stands in contrast to the dynamos of \citet{Augustson15} and \citet{Guerrero16}, in which significant modulation of the DR (10--20\%) played a significant role. For the fourfold-wreath cycle, modulation of mean flows is undetectable. For the partial-wreath cycle, there is a $\sim$$5\%$ weaker DR when compared to the fourfold-wreath cycle (due to the stronger magnetic fields), and a $\sim$$1\%$ modulation of the DR from cycle to cycle. The MC develops significant equatorial asymmetry during partial-wreath cycling, but this appears to be mostly a response to the magnetic field and has little effect on the dynamics. 

\subsection{Dynamics of the Fourfold-Wreath Cycle}
The magnetic field in our simulations is dominated by the azimuthal component $B_\phi$. We therefore first turn our attention to the $\phi$-component of the longitude-averaged Equation \eqref{eq:ind}, i.e.,
\begin{align}
\pderiv{\av{B_\phi}}{t} &\approx \underbrace{(\av{\bm{B}}\cdot\nabla\av{\bm{v}})_\phi}_{{\rm{MS}}_\phi} + \underbrace{\av{\bm{B}^\prime\cdot\nabla\bm{v}^\prime}_\phi}_{{\rm{FS}}_\phi}\nonumber\\
& \underbrace{-\av{\bm{v}\cdot\nabla\bm{B}}_\phi}_{\rm{TA}_\phi} \underbrace{- \av{\nabla\times[\eta(r)\nabla\times\bm{B}]}_\phi}_{{\rm{RD}}_\phi},
\label{eq:dbphidt}
\end{align}
 where ``MS," ``FS," ``TA," and ``RD" refer to mean shear, fluctuating shear, total advection, and resistive dissipation, respectively. The generation due to fluid compression, $-\av{B_\phi\nabla\cdot\bm{v}}$, is very weak and we neglect it throughout. 
 
\begin{figure*}
  	\centering
	\includegraphics[scale=0.9]{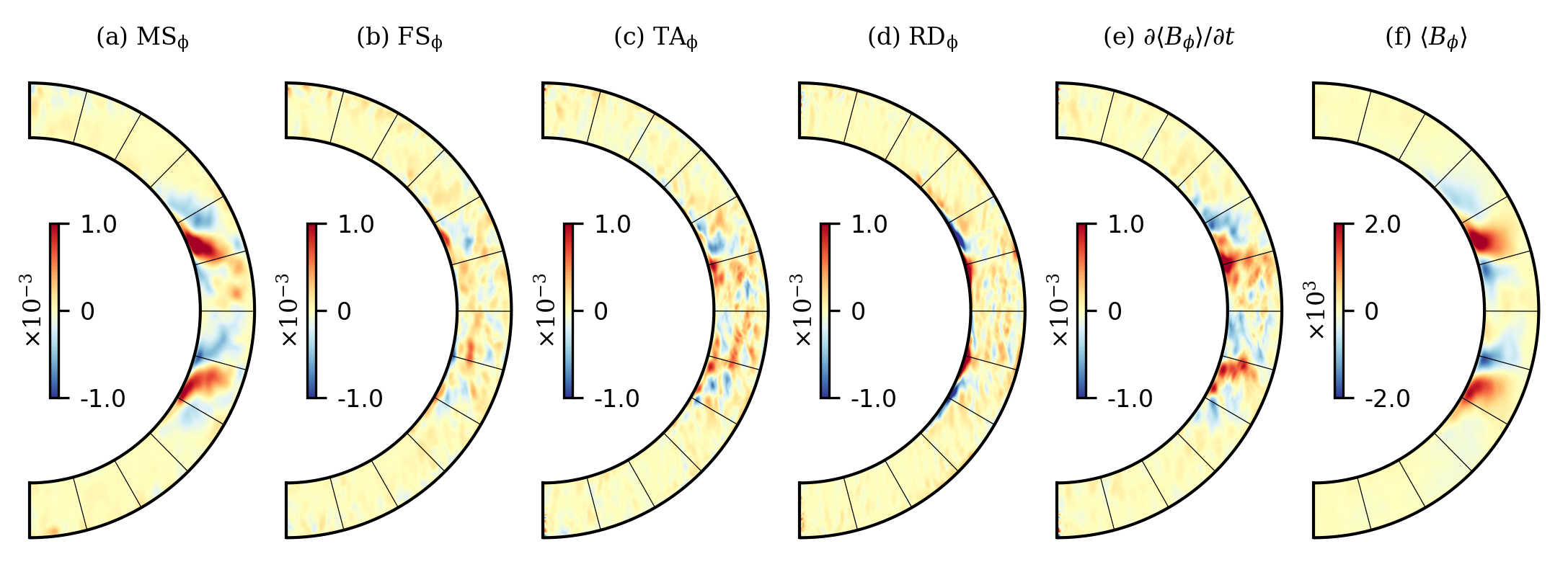}
	\caption{Generation terms in the azimuthal component of the longitude-averaged induction Equation \eqref{eq:dbphidt} for case D3-1, averaged an interval of one rotation period centered at $660\ \prot$. The terms are (\textit{a}) mean shear, (\textit{b}) fluctuating shear, (\textit{c}) total advection, (\textit{d}) resistive dissipation, (\textit{e}) sum of all terms (approximately equal to the time-derivative of $\av{B_\phi}$), and (\textit{f}) $\av{B_\phi}$ itself. The units are c.g.s. ($\rm{G}\ s^{-1}$ in panels (\textit{a})--(\textit{e}) and G in panel (\textit{f})).}
	\label{fig:bphi_generation_sunspot}
\end{figure*}

 Figure \ref{fig:bphi_generation_sunspot} shows profiles of the terms in Equation \eqref{eq:dbphidt} averaged over a short time interval during fourfold-wreath cycling. The time interval chosen is not unique and the profiles in Figure \ref{fig:bphi_generation_sunspot} are representative of the generation terms at any time the fourfold-wreath cycle is dominant.  The governing term is clearly the mean-shear, or $\Omega$-effect ($\rm{MS_\phi}$).  The fluctuating shear ($\rm{FS_\phi}$) is the weakest term overall and has no obvious pattern in its spatial distribution. The total advection ($\rm{TA_\phi}$) is weak and also has no cellular structure (compare to Figure \ref{fig:H3}(\textit{d})), suggesting that transport of $B_\phi$ by the meridional circulation is negligible. The resistive dissipation ($\rm{RD_\phi}$) plays some role (especially near the inner boundary) and has a sign structure opposite to that of $\av{B_\phi}$, acting to destroy whatever field is already present. 
 
 The mean-shear production of $\av{B_\phi}$ has a similar pattern to $\av{B_\phi}$ itself, but with the four wreaths displaced closer to the equator, which tends to drive the $\av{B_\phi}$ wreath structure equatorward. Furthermore, like the fourfold wreaths themselves, the wreaths of $\rm{MS_\phi}$ reappear at mid-latitudes after they disappear near the equator. In Figure \ref{fig:bphi_generation_sunspot}(\textit{a}), for example, the negative-$\rm{MS_\phi}$ wreath straddling ($-15^\circ$)-latitude in the southern hemisphere is associated with a weak negative-$\rm{MS_\phi}$ wreath forming at $\sim$($-40^\circ$)-latitude. Positive $B_\phi$ thus forms at high latitudes where originally there was negative $B_\phi$, creating a polarity reversal. 
 
 It is unclear exactly what stops the equatorward migration---i.e., why the wreaths have significantly reduced amplitude in the $\pm10^\circ$-latitude band straddling the equator. In the Sun, there is presumably only one wreath per hemisphere, each with opposite polarity (as inferred from the solar butterfly diagram). When the two wreaths meet at the equator, they annihilate, as in magnetic reconnection. The same process cannot occur in the fourfold-wreath cycle because the wreaths are often equatorially symmetric. 
 
\begin{figure}
	\includegraphics{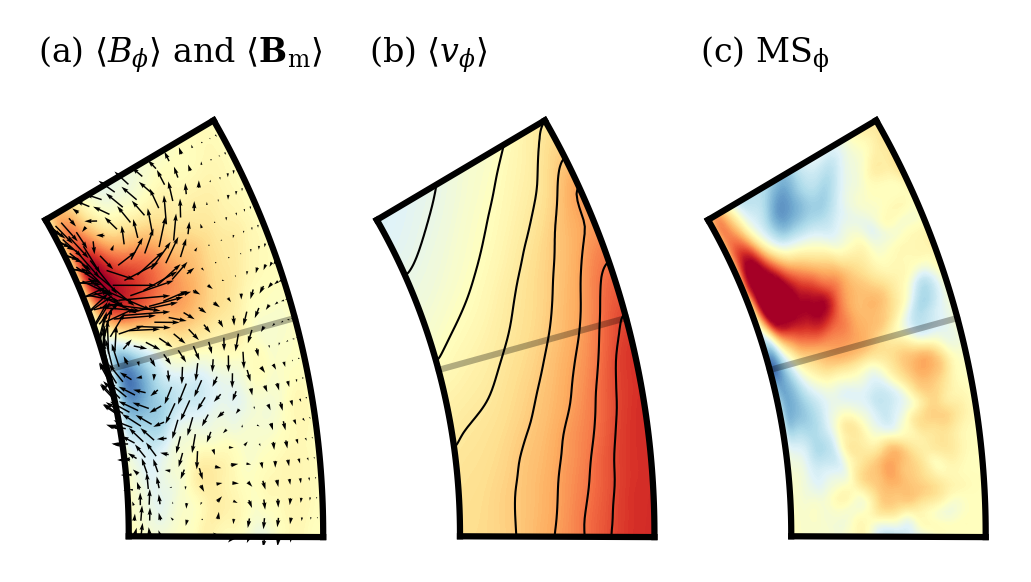}
	\caption{Instantaneous profiles (at $t=660\ \prot$) of (\textit{a}) the longitude-averaged magnetic field, (\textit{b}) the longitude-averaged azimuthal velocity, and (\textit{c}) the mean-shear production $\rm{MS_\phi}$ between the equator and $30^\circ$-north for case D3-1. In (\textit{a}), $\av{B_\phi}$ is plotted in color (with saturation values $\pm2.5$ kG) and $\av{\bm{B}_{\rm{m}}}$ is plotted as a vector field (with the longest arrows corresponding to $\sim$$750$ G). In (\textit{b}) and (\textit{c}), $\av{v_\phi}$ and $\rm{MS_\phi}$ are shown in color with saturation values of $\pm750\ \rm{m\ s^{-1}}$ and $\pm0.001\ \rm{G\ s^{-1}}$, respectively. }
	\label{fig:bm_bphi}
\end{figure}

 We now investigate the origins of the displacement of the shear-production pattern relative to the azimuthal magnetic field. The mean shear is (neglecting curvature terms)
\begin{align}
{\rm{MS}}_\phi \approx \av{\bm{B}_{\rm{m}}}\cdot\nabla \av{v_\phi}.
\label{eq:msphi}
\end{align}
The magnetic field components $\av{\bm{B}_{\rm{m}}}$ and $\av{B_\phi}$, the azimuthal velocity field $\av{v_\phi}$, and the shear production $\rm{MS_\phi}$ are shown as a set of snapshots in Figure \ref{fig:bm_bphi}.  The meridional magnetic field has a fourfold structure and is roughly coincident spatially with the $\av{B_\phi}$ wreaths. From Figure \ref{fig:bm_bphi}(\textit{a}), the streamlines of $\bm{B}_{\rm{m}}$ are roughly counterclockwise for positive $\av{B_\phi}$ and clockwise for negative $\av{B_\phi}$. 

The profile of $\av{v_\phi}$ comes from a differential rotation (at least during the fourfold-wreath cycle) nearly identical to that of the hydrodynamic progenitor, case H3 (compare to Figure \ref{fig:H3}(\textit{c})). The direction field of $\av{\nabla v_\phi}$ (perpendicular to the contours in Figure \ref{fig:bm_bphi}(\textit{b})) is thus cylindrically outward and slightly tilted toward the equator. The combination of $\av{\bm{B}_{\rm{m}}}$ and $\av{\nabla v_\phi}$ creates the pattern of $\rm{MS_\phi}$ shown in Figure \ref{fig:bm_bphi}(\textit{c}).

The overall geometry depicted in Figure \ref{fig:bm_bphi} thus explains the displacement of the term $\rm{MS}_\phi$ relative to $\av{B_\phi}$. In the positive-$\av{B_\phi}$ wreath in the northern hemisphere, for example, the field lines wind so that $\av{\bm{B}_{\rm{m}}}$ points outward in the lower-latitude part of the wreath (corresponding to positive $\rm{MS}_\phi$ via Equation \eqref{eq:msphi}) and inward in the higher-latitude part of the wreath (corresponding to negative $\rm{MS}_\phi$). With time, the positive-$\av{B_\phi}$ wreath is thus strengthened at lower latitudes and weakened at higher latitudes, sending the whole structure south toward the equator so that the positive-$\av{B_\phi}$ wreath is eventually replaced with a negative-$\av{B_\phi}$ wreath. 

  \begin{figure}
	\includegraphics{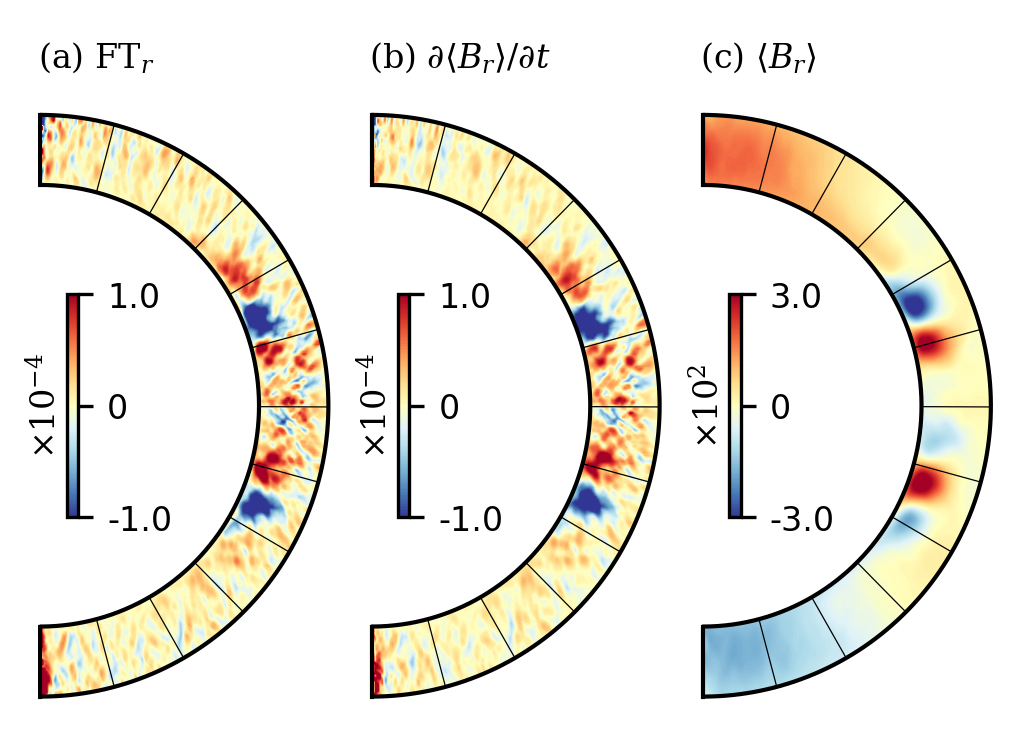}
	\caption{Generation terms appearing in Equation \eqref{eq:dbrdt}, averaged from snapshots taken at the beginning of 57 successive fourfold-wreath cycles. (\textit{a}) Fluctuating transport,  (\textit{b}) sum of all terms (equal to the time-derivative of $\av{B_r}$), and (\textit{c}) $\av{B_r}$ itself. The units are c.g.s. ($\rm{G}\ s^{-1}$ in panels (\textit{a})--(\textit{b}) and G in panel (\textit{c})).}
	\label{fig:br_generation_sunspot}
\end{figure}

The nature of the $\Omega$-effect can also be used to estimate the fourfold-wreath cycle period. Since the mean shear term is responsible for the cycling of $\av{B_\phi}$, the cycle period should be given roughly by 

\begin{align}
P_{\rm{fourfold}} \sim \frac{4\av{B_{\phi}}_{\rm{max}}}{|\rm{MS}_\phi|} \sim \frac{4\av{B_{\phi}}_{\rm{max}}}{|\bm{B}_{\rm{m}}| |\nabla v_\phi|}, 
\end{align}
where $\av{B_{\phi}}_{\rm{max}}$ is the typical azimuthal magnetic field maximum amplitude in the wreath's core and $|\rm{MS_\phi}|$, $|\bm{B}_{\rm{m}}|$ and $|\nabla v_\phi|$ are typical rms-magnitudes of $\rm{MS_\phi}$, $\av{\bm{B}_{\rm{m}}}$, and $\av{\nabla v_\phi}$ (respectively) throughout the whole wreath. For the wreath pair in Figure \ref{fig:bm_bphi} (taking the mean between the equator and 30$^\circ$-north in lower half of the CZ), we compute $\av{B_{\phi}}_{\rm{max}} = 2.63\ \rm{kG}$, $|\bm{B}_{\rm{m}}| = 302\ \rm{G}$, and $|\nabla v_\phi| = 2.17\times10^{-6}\ \rm{s}^{-1}$, yielding
\begin{align}
P_{\rm{fourfold}} \sim 22.0\ \prot, 
\end{align}
in good agreement with the actual cycle period of Equation \eqref{eq:actual_sunspot_period}. 

To investigate the source of the meridional field $\bm{B}_{\rm{m}}$, we turn to the radial component of the longitude-averaged Equation \eqref{eq:ind}, which we write as
\begin{align}
\pderiv{\av{B_r}}{t} &= \underbrace{[\nabla\times(\av{\bm{v}}\times\av{\bm{B}})]_r}_{{\rm{MT}}_r} + \underbrace{[\nabla\times(\av{\bm{v}^\prime\times\bm{B}^\prime})]_r}_{{\rm{FT}}_r} \nonumber\\
& \underbrace{- \av{\nabla\times[\eta(r)\nabla\times\bm{B}]}_r}_{{\rm{RD}}_r}. 
\label{eq:dbrdt}
\end{align}
Here ``MT" and ``FT" refer to the mean transport (magnetic field generation from the mean electromotive force---or e.m.f.---$\av{\bm{v}}\times\av{\bm{B}}$) and the fluctuating transport (or generation from the fluctuating e.m.f. $\av{\bm{v}^\prime\times\bm{B}^\prime}$), respectively. 

Figure \ref{fig:br_generation_sunspot} shows the dominant generation terms for $\av{B_r}$ alongside $\av{B_r}$ itself. We only show the fluctuating transport ${\rm{FT}}_r$, which is an order of magnitude larger than both ${\rm{MT}}_r$ and ${\rm{RD}}_r$. Similar to the generation of $\av{B_\phi}$ shown in Figure \ref{fig:bp_during_sunspot_cycle}, the profile for $\partial \av{B_r}/\partial t$ is displaced equatorward relative to $\av{B_r}$, leading to the equatorward propagation of $\av{B_r}$ as well as $\av{B_\phi}$. The generation terms for $\av{B_\theta}$ (not shown) possess a similar structure, with ${\rm{FT}}_\theta$ dominating the other terms by a factor of $\sim$5. 

The fluctuating transport is similar in spirit to the $\alpha$-effect from mean-field theory, which represents the e.m.f. caused by the helical action of convection on the magnetic field. In this respect, the fourfold-wreath cycle is characteristic of an $\alpha$--$\Omega$ dynamo. However, our simulations are distinct from most mean-field dynamo models in that the generation of both meridional field and azimuthal field occurs in the same location in the CZ, with the meridional circulation playing essentially no role. This stands in contrast to the flux-transport paradigm, in which the $\alpha$-effect creates meridional field at high latitudes that is then pumped deep into the CZ by the meridional circulation to seed the $\Omega$-effect in the following cycle. 

\subsection{Dynamics of the partial-wreath cycle}
The physical mechanisms responsible for the system transitioning between the fourfold-wreath and partial-wreath states remain unclear. Nevertheless, we can understand the maintenance of the dynamo during the partial-wreath state in terms of the dominant induction terms. Furthermore, even though the partial-wreath structure is inherently non-axisymmetric, it is sufficient to perform a mean-field analysis---i.e., identify the dominant \textit{longitude-averaged} induction terms of Equations \eqref{eq:dbphidt} and \eqref{eq:dbrdt}. Since the longitude-averaged field $\av{B_\phi}$ has the same polarity as the dominant partial wreath, there is a one-to-one correspondence between cycles in $\av{B_\phi}$ and partial-wreath cycles. 

 
 
\begin{figure*}
	\includegraphics{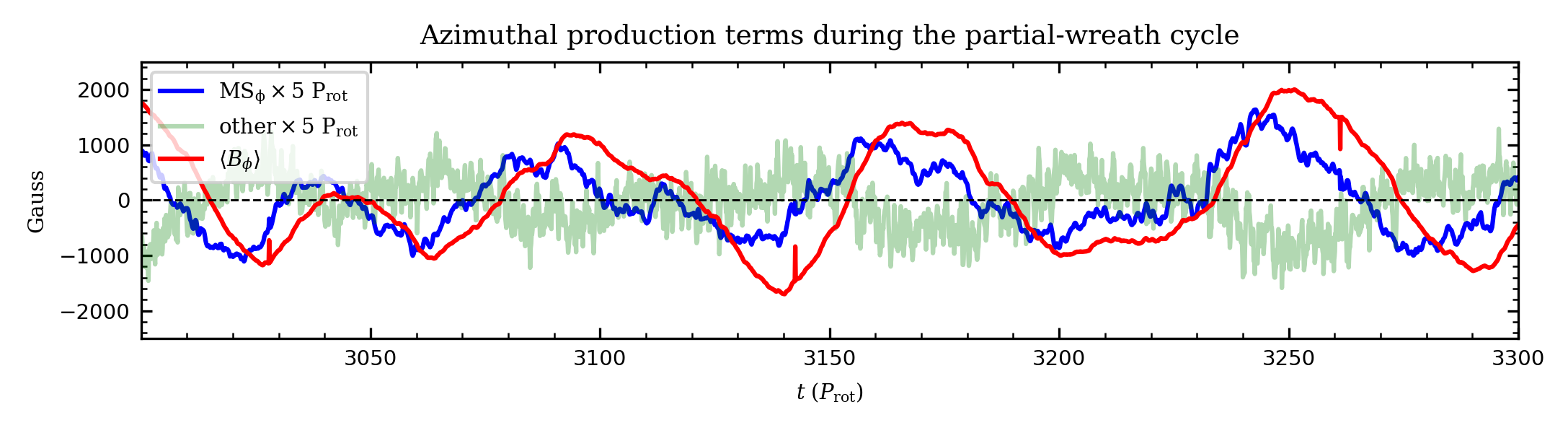}
	\caption{The dominant generation terms appearing in Equation \eqref{eq:dbrdt}, as well as $\av{B_\phi}$, when the partial-wreath structure is cycling in the south, during the same interval shown in Figure \ref{fig:zoom_tl_bphi_mid}). Quantities are averaged over radius and between $0^\circ$ and $20^\circ$-south in latitude. The generation terms are broken up into mean-shear and ``other" (i.e., $\partial\av{B_\phi}/\partial t - \rm{MS_\phi}$) and multiplied by 5 $\prot$ to put them on the same scale as $\av{B_\phi}$. The contribution to ``other" dominated by the resistive dissipation. }
	\label{fig:bphi_generation_asymmetric}
\end{figure*}

The mean-field generation terms during partial-wreath cycling vary rapidly in space and time. Therefore, significant spatial and temporal averaging must be performed to extract a meaningful signal. Figure \ref{fig:bphi_generation_asymmetric} shows the spatially-averaged contributions to the production of $\av{B_\phi}$ in case D3-1 from the mean shear (which is clearly dominant) and the other terms, as well as $\av{B_\phi}$ itself, during the interval of partial-wreath cycling depicted in Figure \ref{fig:zoom_tl_bphi_mid}. All three profiles are roughly sinusoidal. The profile of mean-shear production slightly leads the profile of $\av{B_\phi}$ in time, indicating that the rise and fall of $\av{B_\phi}$ is directly caused by the mean-shear production. The contribution from the other induction terms (which is dominated by the resistive  dissipation) is $\sim$$180^\circ$ out-of-phase with $\av{B_\phi}$, meaning the magnetic diffusion attempts to destroy whatever field is already there. 
 
The production of $\av{\bm{B}_{\rm{m}}}$ (which in turn affects the production of $\av{B_\phi}$ through mean-shear) is highly chaotic in time during the partial-wreath cycle, and so instead of plotting the production terms in Equation \eqref{eq:dbrdt} as functions of time, we simply quote their temporally-averaged magnitudes below. Just as in the fourfold-wreath cycle, the dominant term producing $\av{\bm{B}_{\rm{m}}}$ is the fluctuating e.m.f.:
\begin{align}
(\rm{FT_r})_{\rm{rms}} &= 9.3\ (\rm{MT_r})_{\rm{rms}}\\
(\rm{FT_r})_{\rm{rms}} &= 2.6\ (\rm{RD_r})_{\rm{rms}}
\end{align}
at mid-depth, where the mean is taken over the time interval (3000, 3300) $\prot$ and the latitude band between $0^\circ$ and $20^\circ$-south. Similar relationships hold for the production of $\av{B_\theta}$, which is also dominated by the fluctuating e.m.f.

\section{Bistability in the Context of Solar Observations}\label{sec:obs}
Continuous, full-disc observations of the photospheric magnetic field, produced by MDI/\textit{SOHO} and HMI/\textit{SDO},  have greatly constrained certain features of the solar dynamo and yielded a number of surprises. \citet{Stenflo12} analyzed a large set of MDI magnetograms from 1995--2011 and found that Joy's law (which states that emerging sunspot pairs are tilted on average such that the leading spot is closer to the equator, and that the tilt is proportional to heliographic latitude) holds for active regions both large and small. Furthermore, there was no observed dependence of tilt angle on total active-region flux (or region area), implying that the tilts are inherent to buoyant flux ropes in the solar interior. This stands in contrast to the argument that the tilts are established by the Coriolis force as the flux ropes rise through the CZ \citep{DSilva93, Fan94}. 

The view that the interior flux ropes are associated with a preferred sense of tilt is interesting in light of the fourfold-wreath cycle. Each of the fourfold wreaths possesses a particular sense of twist as in Figure \ref{fig:bm_bphi}(\textit{a}). If loops were to spontaneously break off from the wreaths (this does not occur in our simulations because the magnetic diffusivity is too high), they would be endowed with a tilt from the twisting magnetic field lines in the wreath. \citet{Nelson14} achieved the separation of magnetic loops from interior wreaths and found that the sample of loops reaching the surface made contact with Joy's law. This supports the idea that a fourfold-wreath-like structure in the Sun could establish systematic tilts associated with flux ropes in the deep interior. 

\citet{Stenflo12} further reported that  a statistically significant fraction of active regions ($5\%$--$25\%$, depending on region size) violate Hale's polarity law, strongly suggesting that there must be multiple reservoirs of interior magnetism from which buoyant flux ropes originate. Additionally, the ``anti-Hale" active regions could appear in the same latitude band as ``Hale" active regions, suggesting close proximity in latitude for the reservoirs. This result agrees with the geometry of both the fourfold-wreath structure (whose wreaths come in opposite-polarity pairs with wreaths in a pair separated in latitude by $\sim$$15^\circ$) and the partial-wreath structure (with opposite-polarity partial wreaths at approximately the same latitude). 

\citet{Li18} performed another extensive analysis of sunspot groups from 1996--2018 using both the MDI and HMI magnetograms, also reporting that a sizeable fraction ($\sim$8\%) of sunspot pairs violate Hale's polarity law and confirming the result obtained by \citet{Stenflo12}. In addition, \citet{Li18} found that the anti-Hale sunspots had different properties than Hale sunspots, namely smaller bipole separation and weaker total magnetic flux. These results further supports the idea that the Hale and anti-Hale sunspots may come from different interior reservoirs. 

It is also possible that the anti-Hale active regions form an essential part of the solar dynamo. \citet{Hathaway16} find that modeling the flux transport of active regions observed by \textit{SDO} during the current solar cycle can be used to predict the amplitude and hemispheric asymmetry of the following cycle. Central to such predictions is the initial distribution of active region tilts, which would be significantly modified by the presence of anti-Hale active regions. Although it is unlikely that the real interior solar magnetic field matches the fourfold-wreath and partial-wreath structures in detail, such configurations offer two ways in which violations of Hale's polarity law could occur. 

We now discuss the asymmetric features of solar activity, many of which are captured by the partial-wreath cycle. For well over a century, sunspots have been observed to emerge preferentially at particular solar longitudes (e.g., \citealt{Maunder1905}; \citealt{Svalgaard75}; \citealt{Bogart82}; \citealt{Henney02}). This phenomenon, referred to as \textit{active longitudes}, remains poorly understood. 

It is common for active longitudes to come in pairs, separated by about 180$^\circ$ in longitude (e.g., \citealt{Bai03}). The pairs are also usually found in a single hemisphere and are part of the global \textit{hemispheric asymmetry} in magnetic activity. Finally, the pairs are associated with opposite polarity of the azimuthal magnetic field on opposite sides of the Sun, as revealed by synoptic maps from the Wilcox Solar Observatory \citep{Mordinov04}. Thus, it is possible that the pairs are associated with nests of Hale and anti-Hale active regions on opposite sides of the Sun, although a systematic study of the MDI/HMI magnetograms would be needed to confirm this result. 

Active longitudes and their associated hemispheric asymmetry bear a striking resemblance to the partial-wreath structure in our simulated dynamos. The partial wreaths preferentially occupy one hemisphere at a time. They come in opposite-polarity pairs with the strongest magnetism on opposite sides of the sphere (see the Mollweide projection in Figure \ref{fig:bphi_twodepths}(\textit{b}) and the 3D volume-renderings in Figure \ref{fig:volrender}). The time-longitude diagrams for the partial wreaths in case D3-1 (Figure \ref{fig:tlon}) show that each wreath is roughly stationary once an appropriate rotating frame is chosen, similar to the quasi-rigid structure formed by active-longitude pairs in the Carrington reference frame \citep{Berdyugina03}.

 Figures \ref{fig:volrender} and \ref{fig:tlon} would also indicate cyclic behavior in which the strongest active longitude flips by 180$^\circ$ over the partial-wreath cycle period. \citet{Berdyugina03} reported such behavior in observations of solar sunspot number, wherein the active longitudes appeared to flip between opposite sides of the Sun every $\sim$3.7 years, or every $\sim$53 solar Carrington rotations. Our simulation suite displays similar properties in the relative energy content of the partial wreaths (Figure \ref{fig:partial_wreath_time_trace}), showing that the dominant partial wreath switches with a quasi-regular period of 81.5 $\prot$.
 
The hemispheric asymmetry in solar activity can be seen in a wide variety of activity indicators, such as number of flares, sunspot number, sunspot area, and photospheric magnetic flux (e.g., \citealt{Schussler18} and references therein). Many studies have been performed on the periodic behavior of the asymmetry, with most agreeing on a long-term period of $\sim$50 years and a short-term period of $\sim$10 years (see \citealt{Hathaway15} and reference therein). Our simulations each possess a significant hemispheric asymmetry (Figure \ref{fig:hemispheric_asymmetry}(\textit{a})), although there is no obvious associated cycle (Figure \ref{fig:hemispheric_asymmetry}(\textit{b})). Nevertheless, our simulations make contact with observations in that the hemispheric asymmetry occurs as a long-term modulation of another cycle, which in our simulations is that of the partial wreaths.



In summary, the dynamo cases presented here show similarities to several features of solar activity, namely the sunspot cycle, opposite-polarity magnetic regions in close proximity, active longitudes, and hemispheric asymmetry. We do not suggest that all these observed features of the Sun's activity are a result of fourfold-wreath and partial-wreath structures existing in the solar CZ. Rather, our simulations provide insights as to how a solar-like convection zone can support dynamos exhibiting a wide range of cyclic behavior.

\section{Conclusions}\label{sec:concl}
We have explored three dynamo simulations of a 3D solar-like convection zone at varying magnetic Prandtl number. These dynamo cases display clear signs of bistability. In particular, there is a fourfold-wreath structure that exhibits equatorward propagation and regular polarity reversals. There is also a uni-hemispherical partial-wreath pair that is roughly stationary and whose whose dominant polarity changes cyclically, but with an irregular period. The fourfold-wreath cycle captures aspects of the 22-year sunspot cycle (namely, the solar butterfly diagram), whereas the partial-wreath cycle is reminiscent of several of the observed features of active longitudes and hemispheric asymmetry. 

We have examined the dynamics of the fourfold-wreath cycle in detail, and find they are approximately described by an $\alpha$-$\Omega$ dynamo. The mean shear, or $\Omega$-effect, is responsible for the generating the wreaths. The regular polarity reversals and equatorward propagation are caused by the inherent geometry of the wreaths (namely the meridional field winding counterclockwise for positive $B_\phi$ and clockwise for negative $B_\phi$) coupled to a solar-like differential rotation. Generation of the meridional field occurs through the fluctuating e.m.f., which is similar to an $\alpha$-effect, but fully nonlinear. Both effects occur in the same localized region of the CZ (with the meridional circulation playing basically no role), unlike in the flux-transport dynamo framework. As a whole, the production terms for the fourfold-wreath cycle are similar to those of model K3S in \citet{Augustson15}, although in that work, significant modulation of the differential rotation played a large role, whereas in our dynamo cases, the differential rotation stays mostly constant with time. 

The partial-wreath cycle has longitude-averaged dynamics similar to an $\alpha$-$\Omega$ dynamo as well, but it is not fully clear how the system transitions between the regularly cycling fourfold-wreaths and the uni-hemispherical partial-wreath pair.  It is possible that the transition is governed by the fourfold-wreath structure changing between equatorially symmetric and antisymmetric states, but more work needs to be done to assess in detail how such a mechanism operates. Equatorial parity also played a key role in model K3S of \citet{Augustson15}, wherein atypical dominance of the radial magnetic field by the even-$\ell$ modes acted as a precursor to the cycling dynamo entering a ``grand minimum." Similarly, in the global convective dynamo simulation of \citet{Raynaud16}, the interaction of hydromagnetic modes with different symmetries led to transitions between distinct cycling dynamo states. The magnetic Prandtl number also may be significant in determining whether the system is unstable to the partial-wreath state; \citet{Nelson13} found magnetic structures resembling uni-hemispherical partial-wreaths in their high-$\prm$ simulations but not low-$\prm$. 

Our work suggests that spherical-shell convection can support complex dynamos capable of exhibiting vastly different global structures for the magnetism simultaneously. In particular, the bistable properties of our dynamo cases are consistent with a solar dynamo containing two cycles at once: one associated with the observed butterfly diagram and one associated with active longitudes and the hemispheric asymmetry. We recognize that our simulations here are just starting to explore the complexities that might be at work in the highly nonlinear and turbulent dynamics of the Sun. The range of scales that we can capture and the diffusivities we use are still far removed from the solar regime. However, we have found that some reasonable large-scale behavior emerges from our dynamo models, and these may broaden our insights into the physical processes occurring deep in the interior of the Sun. 


\begin{appendices}
\section*{Appendix}\nonumber
\section{Definitions of non-dimensional numbers and diffusion times}\label{ap:nond}
After the example of \citet{Featherstone16b}, we parameterize the strength of the imposed driving through a bulk ``flux Rayleigh number,"
\begin{align}
\raf \define \frac{\tilde{g}\tilde{F}H^4}{\cp\tilde{\rho}\tilde{T}\nu^3}, 
\label{def:raf}
\end{align}
where $H$  is the shell depth and $F$ is the energy flux associated with the combined conduction and convection in a statistically steady state. The tildes refer to volume-averages over the full spherical shell. 

We parameterize the degree of rotational constraint through a bulk Rossby number,
\begin{align}
{\rm{Ro}} \define \widetilde{\bigg{(}\frac{v^\prime}{2\Omega_0 H_v}\bigg{)}},
\label{def:ro}
\end{align}
where the typical velocity $v^\prime(r)$ refers to the rms of the convective ($|m|>0$) velocity over a sphere of radius $r$.  The typical length-scale of the flow structure $H_v(r)$ is not well-represented by either the scale height or the shell depth. Instead, we choose $H_v(r)$ based on the power spectrum of the convective velocity. Specifically, $H_v(r) \define 2\pi r/\ell_v(r)$, where $\ell_v(r) \define \sum_{|m|>0}\ell |\hat{\bm{v}}_{\ell m}|^2$ and $|\hat{\bm{v}}_{\ell m}|^2$ refers to the normalized $(\ell,m)$-power associated with $\bm{v}$ at radius $r$. We similarly define define a spectral length-scale $H_B(r)$ associated with the convective magnetic field structures.

We assess the level of turbulence through a volume-averaged Reynolds number,
\begin{align}
{\rm{Re}} \equiv \widetilde{\bigg{(}\frac{v^\prime H_v}{\nu}\bigg{)}}
\label{def:re}
\end{align}
and a volume-averaged magnetic Reynolds number,
\begin{align}
{\rm{Re_m}} \equiv \widetilde{\bigg{(}\frac{v^\prime H_B}{\eta}\bigg{)}},
\label{def:rem}
\end{align}

We define the thermal diffusion time (which is equal to the viscous diffusion time since Pr = 1) as 
\begin{align}
\rm{TDT} &= \frac{(\ro - \ri)^2}{\kappa(\ro)} = 115\ \prot,
\end{align}
a constant for all four cases. 

We define the magnetic diffusion time as 
\begin{align}
\rm{MDT} &= \frac{(\ro - \ri)^2}{\eta(\ro)} = 115,\ 230,\ 460\ \prot
\end{align}
for cases D3-1, D3-2, and D3-4, respectively. 
\end{appendices}

\acknowledgments
The authors thank Sacha Brun, Kyle Augustson, Connor Bice, and Bradley Hindman for helpful conversations surrounding (among other topics) convection-driven spherical-shell dynamos. We appreciate the detailed suggestions of the reviewer, which we believe significantly improved the manuscript. L.I.M. was partly supported during this work by the Future Investigators in NASA Earth and Space Sciences Technology (FINESST) award 80NSSC19K1428 and by a George Ellery Hale Graduate Fellowship. This research was primarily supported by NASA Heliophysics through grants 80NSSC18K1127, NNX17AG22G, and NNX13AG18G. Computational resources were provided by the NASA High-End Computing (HEC) Program through the NASA Advanced Supercomputing (NAS) Division at Ames Research Center. Rayleigh has been developed by Nicholas Featherstone with support by the National Science Foundation through the Computational Infrastructure for Geodynamics (CIG) under NSF grants NSF-0949446 and NSF-1550901.


\clearpage

\end{document}